\documentclass[twocolumn]{aastex63}

\usepackage{lineno}
\usepackage[normalem]{ulem}
\usepackage{booktabs}
\usepackage{float} 
\usepackage{balance}
\usepackage{cancel}

\newcommand\kms{{\rm\,km\,s^{-1}}}

 \def\mso{\,\mathrm{M}_\odot}

 \def\simle{\mathrel{\hbox{\rlap{\hbox{\lower4pt\hbox{$\sim$}}}\hbox{$<$}}}}
 \def\simgr{\mathrel{\hbox{\rlap{\hbox{\lower4pt\hbox{$\sim$}}}\hbox{$>$}}}}
\renewcommand{\thefigure}{\arabic{figure}}

\received{2025 January 14}
\revised{2025 February 6}
\accepted{2025 February 14}
\submitjournal{ApJL}

\shorttitle{Red stragglers}
\shortauthors{Chen Wang et al.}

\begin{document}
%\linenumbers

\title{Using detailed single star and binary evolution models to probe the large observed luminosity spread of red supergiants in young open star clusters}

\correspondingauthor{Chen Wang}
\email{cwang@mpa-garching.mpg.de}

\author[0000-0002-0716-3801]{Chen Wang}
\affiliation{Max Planck Institute for Astrophysics, Karl-Schwarzschild-Strasse 1, 85748 Garching, Germany}

\author[0000-0002-9015-0269]{Lee Patrick}
\affiliation{Departamento de Astrof\'{i}sica, Centro de Astrobiolog\'{i}a, (CSIC-INTA), Ctra. Torrej\'{o}n a Ajalvir, km 4, 28850 Torrej\'{o}n de Ardoz, Madrid, Spain}
%\affiliation{Departamento de F\'{i}sica Aplicada, Universidad de Alicante, E-03690 San Vicente del Raspeig, Alicante, Spain}
%\affiliation{School of Physical Sciences, The Open University, Walton Hall, Milton Keynes MK7 6AA, UK}

\author[0000-0002-2715-7484]{Abel Schootemeijer}
\affiliation{Argelander-Institut f\"ur Astronomie, Universit\"at Bonn, Auf dem H\"ugel 71, 53121 Bonn, Germany}

\author[0000-0001-9336-2825]{Selma E. de Mink}
\affiliation{Max Planck Institute for Astrophysics, Karl-Schwarzschild-Strasse 1, 85748 Garching, Germany}

\author[0000-0003-3026-0367]{Norbert Langer}
\affiliation{Argelander-Institut f\"ur Astronomie, Universit\"at Bonn, Auf dem H\"ugel 71, 53121 Bonn, Germany}

\author[0000-0003-3996-0175]{Nikolay Britavskiy}
\affiliation{Universit\'{e} de Li\`{e}ge, Quartier Agora (B5c, Institut d’Astrophysique et de G\'{e}ophysique), All\'{e}e du 6 Ao\v{u}t 19c, 4000 Sart Tilman, Li\`{e}ge, Belgium}
\affiliation{ Royal Observatory of Belgium, Avenue Circulaire/Ringlaan 3, 1180 Brussels, Belgium}

\author[0000-0001-9565-9462]{Xiao-Tian Xu}
\affiliation{Argelander-Institut f\"ur Astronomie, Universit\"at Bonn, Auf dem H\"ugel 71, 53121 Bonn, Germany}

\author[0000-0002-9552-7010]{Julia Bodensteiner}
\affiliation{European Southern Observatory, Karl-Schwarzschild-Str. 2 85738 Garching bei München, Germany}
\affiliation{Anton Pannekoek Institute for Astronomy, University of Amsterdam, Postbus 94249, 1090GE Amsterdam, Nederland}

\author[0000-0003-1009-5691]{Eva Laplace}
\affiliation{Heidelberger Institut für Theoretische Studien, Schloss-Wolfsbrunnenweg 35, D-69118 Heidelberg, Germany}

\author[0000-0003-3456-3349]{Ruggero Valli}
\affiliation{Max Planck Institute for Astrophysics, Karl-Schwarzschild-Strasse 1, 85748 Garching, Germany}

\author[0000-0003-1817-3586]{Alejandro Vigna-G\'omez}
\affiliation{Max Planck Institute for Astrophysics, Karl-Schwarzschild-Strasse 1, 85748 Garching, Germany}

\author[0000-0002-7527-5741]{Jakub Klencki}
\affiliation{European Southern Observatory, Karl-Schwarzschild-Str. 2 85738 Garching bei München, Germany}
\affiliation{Max Planck Institute for Astrophysics, Karl-Schwarzschild-Strasse 1, 85748 Garching, Germany}

\author[0000-0001-7969-1569]{Stephen Justham}
\affiliation{Max Planck Institute for Astrophysics, Karl-Schwarzschild-Strasse 1, 85748 Garching, Germany}

\author[0000-0002-3054-4135]{Cole Johnston}
\affiliation{Max Planck Institute for Astrophysics, Karl-Schwarzschild-Strasse 1, 85748 Garching, Germany}

\author[0000-0002-9911-8767]{Jing-ze Ma}
\affiliation{Max Planck Institute for Astrophysics, Karl-Schwarzschild-Strasse 1, 85748 Garching, Germany}
%===================================================================================================
\begin{abstract}
%No more than 250 workds
%===================================================================================================

Red supergiants (RSGs) represent a late evolutionary stage of massive stars. Recent observations reveal that the observed luminosity range of RSGs in young open clusters is wider than expected from single star evolution models. Binary evolution effects have been suggested as a possible explanation. Here, we analyse 3670 detailed binary-evolution models, as well as corresponding single-star models, to probe the contribution of binary mass transfer and binary mergers on the luminosity distribution of RSGs in star clusters with ages up to 100 Myr. We confirm that the expected luminosity range of RSGs in a coeval population can span a factor of ten, as a consequence of mergers between two main-sequence stars, which reproduces the observed red supergiant luminosity ranges in rich clusters well. While the luminosity increase as consequence of mass transfer is more limited, it may help to increase the number of overluminous RSGs.
However, our results also demonstrate that binary effects alone are insufficient to account for the number of RSGs found with luminosities of up to three times those predicted by current single-star models. We discuss observational accuracy, rotational mixing, age spread, and intrinsic RSG variability as possible explanations. Further observations of RSGs in young open clusters, in particular studies of their intrinsic brightness variability, appear crucial for disentangling these effects.

\end{abstract}

%===================================================================================================
\section{Introduction} \label{sec:intro}
%No more than 3500 words, no more than 5 combined figures (each limited to 9 panels), including figures and tables
%===================================================================================================
Red supergiants (RSGs) are cool giant massive stars that represent a late phase of stellar evolution, where the star has exhausted hydrogen in its core and is now fusing heavier elements. They are considered progenitors of hydrogen-rich Type II-P supernovae (SN). 
Understanding the formation and evolution of RSGs is crucial for advancing our knowledge of stellar evolution and the chemical yields from SN explosions. Despite their importance, key processes governing RSG formation and evolution, such as the roles of stellar rotation and binary interaction, remain poorly constrained.

It is widely accepted that massive stars are primarily born with close companions \citep{2012Sci...337..444S,2023ASPC..534..275O}. Studies have shown that binary evolution significantly alters the luminosity of RSGs \citep{2008MNRAS.384.1109E,2020MNRAS.495L.102E,2019A&A...631A...5Z,2024A&A...686A..45S}. Young open clusters provide an excellent laboratory for studying these effects, as their coeval stellar populations would produce similar luminosities among RSGs if only single-star evolution were at play.
Using BPASS single and binary models, \cite{2020MNRAS.495L.102E} demonstrated that binary evolution can greatly extend the luminosity distribution of RSGs in young star clusters. Mergers and mass accretors can appear more luminous than other RSGs, whereas RSGs undergoing mass transfer may have lower luminosities. 
In fact, recent high-quality spectroscopic observations reveal that the luminosity distribution of RSGs in young open clusters is too broad to be explained by single-star evolution \citep{2019A&A...624A.128B,2020A&A...635A..29P}. The more luminous RSGs, known as “red stragglers”, appear younger and more massive than their counterparts and are thought to be results of binary evolution.

From another point of view, based on the analysis of large amounts of detailed binary models, our previous studies \citep{2020ApJ...888L..12W,2022NatAs...6..480W} show that binary evolution is crucial in explaining the extended main-sequence (MS) turn-off in young open clusters observed by the Hubble Space Telescope (HST) over the past decade \citep{2009A&A...497..755M,2015MNRAS.450.3750M,2018MNRAS.477.2640M,2023A&A...672A.161M,2017ApJ...844..119L}. We proposed that MS mergers (mergers between two MS stars) produce hot and luminous blue stragglers, whereas stable mass transfer mainly results in fast-rotating Be stars that are cooler than other MS stars in the cluster. These MS stars are expected to exhibit distinct properties during later stages, such as the RSG phase. For instance, blue stragglers are expected to evolve into red stragglers as they progress through their evolution.

Previous studies have suggested that RSGs are unique indicators of cluster age \citep{2019MNRAS.486..266B,2019A&A...624A.128B,2020MNRAS.495L.102E}, especially given the fact that the MS turn off is usually extended in young open clusters \citep{2018MNRAS.477.2640M,2023A&A...672A.161M}. While some studies argue that the least luminous RSGs are ideal age indicators since more luminous red stragglers are likely binary products \citep{2019MNRAS.486..266B,2019A&A...624A.128B}, others suggest that the mean luminosity of all RSGs provides a better measure, as this value is similar in both single and binary evolution \citep{2020MNRAS.495L.102E}. Thus, understanding the influence of binary interaction on RSGs is essential for accurately using them as age indicators.

In this study, we investigate the RSG phase of our detailed binary models from \cite{2022NatAs...6..480W}, offering the first comprehensive analysis of RSGs in young open clusters based on a large set of detailed binary models. 
Compared to the BPASS models used in \cite{2020MNRAS.495L.102E}, our models evolve both stars simultaneously and include the spin-up of accretors. Our models also treat mass transfer stability and efficiency in a self-consistent approach. Furthermore, compared to \cite{2020MNRAS.495L.102E} which focuses solely on the luminosity spread of RSGs without considering the time they spend at each luminosity, our study provides a comprehensive population synthesis analysis of the luminosity distribution and number of RSGs from various evolutionary pathways.

We compare our results with observed RSGs in four young open clusters, NGC\,330 \citep{2020A&A...635A..29P} in the Small Magellanic Cloud (SMC) and NGC\,2004, NGC\,2100 and NGC\,1818 (\citealt{2016MNRAS.463.1269B,2019MNRAS.486..266B} and N. Britavskiy in private communication) in the Large Magellanic Cloud (LMC). We aim to enhance our understanding of the properties of RSGs originating from different pathways and provide useful constraints on massive binary evolution. We select to study these four clusters because they are in Magellanic Clouds - thus,  from the observational point of view there is less extinction and as a result, the luminosity estimates are more reliable for RSGs in these clusters (the luminosity estimates are mainly from the K-band photometry). While other clusters with RSGs are known in the Galaxy \citep[see Table\,1 in][]{2020MNRAS.495L.102E}, 
their luminosity estimates (and spread) are less robust 
due to uncertain extinction around RSGs, especially in very crowded regions in the Galaxy. Therefore, studying RSGs in Magellanic Cloud clusters provides a more secure basis for our analysis than studying those in the Galaxy.

The \textit{Letter} is structured as follows: in Section\,\ref{sec:method} we introduce our detailed single and binary models. In Section\,\ref{sec:Results}, we demonstrate the predicted properties of RSGs in young open clusters from both single star and binary evolution and compare with recent observations. Then we discuss the impact of uncertainties on our results in Seciont\,\ref{sec:discussion} and summarize our conclusions in Section\,\ref{sec:conclusions}.

%===================================================================================================
\section{Models and physics assumptions}\label{sec:method}
%===================================================================================================

\subsection{Detailed single-star models}\label{sec:single_model}
We first explore whether the observed RSGs in four young open clusters can be explained by single-star evolution. For this purpose, we employ single-star models computed by \citet{2021A&A...653A.144H} using the MESA code (version 12115, \citealt{Paxton2011,Paxton2013,Paxton2015,2018ApJS..234...34P}).
These models cover initial masses ranging from 2 to 20$\mso$. We extended these models up to initial masses of 100$\mso$, with a mass interval of $\log m_{\rm i} = 0.04$. These models adopt physics assumptions nearly identical to those in \citet{2019A&A...625A.132S}, except for a mass-dependent overshooting parameter. Specifically, $\alpha_{\rm OV} = 0.3$ is used for stars with initial masses above $20 \mso$ \citep{2011A&A...530A.115B}. Below $20 \mso$, $\alpha_{\rm OV}$ decreases linearly, reaching 0.1 at an initial mass of $1.66 \mso$ \citep{2016A&A...592A..15C}. Semiconvective mixing is set to $\alpha_{\rm SC} = 10$ \citep{2019A&A...625A.132S}.
The evolution starts from the zero-age main sequence (ZAMS) and terminates when core helium is depleted. We identify RSGs as stellar models with $T_\mathrm{eff}<4800\,\mathrm K$ \citep{2009ApJ...703..441D}.

We consider both LMC and SMC metallicities. For direct comparison with observed RSGs (in Figs.\,\ref{fig:n330_single_track} and \ref{appfig:3cluster_single_track}), we use non-rotating star models. Additionally, we employ rotating star models with the same physics assumptions for two parts in this work. 
First, we use stellar models initially rotating at 15\% of their critical values to follow the evolution of MS merger products in our detailed binary models. Second, in our semi-analytical method to calculate the upper luminosity limit of RSGs in young open clusters (see Sections\,\ref{sec:RSG_MS_Merger} and Appendix\,\ref{app_sec:semi_analytical} for detail), we employ single-star models with an initial rotation of 55\% of their critical velocity for pre-merger stars. These two rotational velocities are chosen based on the findings of \cite{2022NatAs...6..480W}, which suggest that most stars are likely born with around 55\% of their critical rotational velocity, contributing to the main (red) MS in young open clusters. Binary merger products, if rotating at approximately 15\% of their critical velocity, can explain the blue MS observed in these clusters. 

\subsection{Detailed binary models}\label{sec:method_binary}
We use the MESA detailed binary models \footnote{We use MESA version 8845 to compute the binary models. The inlist files used for these computations can be found at \dataset[Zenodo]{https://zenodo.org/records/5233209}} with the SMC metallicity in \citet{2022NatAs...6..480W} and \citet{2024ApJ...975L..20W}. The initial binary parameters are created from Monte Carlo sampling. The initial primary masses are from 3 to 100$\mso$ adopting the \cite{Salpeter1955} initial mass function (IMF). The initial mass ratios range from 0.1 to 1, following a flat distribution \citep{2012Sci...337..444S,2022A&A...665A.148S}. The initial orbital periods are from the minimum value, where the two stellar components are initially in contact, to 8.6\,yrs ($\log P_{\rm i}\, /\rm d=3.5$), assuming a flat distribution in logarithmic space \citep{1924PTarO..25f...1O,1983ARA&A..21..343A,2023A&A...674A..60B}. We simulate the evolution of 3670 detailed binary systems, representing clusters with an initial total stellar mass of approximately $1.3 \times 10^5 \mso$, under the assumption of a 100\% binary fraction and a lower mass limit of $0.08\mso$ (see \cite{2024ApJ...975L..20W} for details).

The adopted upper period limit of $\log P=3.5$ reliably reaches the non-interaction region for binaries with primary masses lower than approximately 25$\mso$ (whose MS lifetime is around 7.6\,Myr). However, the true upper period limit for binaries remains unclear, especially in dense environments such as star clusters, where dynamical evolution is expected to disrupt wide systems \citep{2012A&A...543A.126K}. For the lower period limit, we use the period at which the two stars initially in contact at ZAMS. Binaries that are not initially in contact but where the primary star fills its Roche lobe at the ZAMS phase are classified as ZAMS mergers. We assume that ZAMS merger products are also ZAMS stars. The fraction of ZAMS mergers, as well as similar populations of stars including pre-MS mergers and mergers occurring before and shortly after the ZAMS phase, remains uncertain. \cite{2022NatAs...6..480W} proposed that a substantial fraction of binaries (between 10\% to 30\%) must merge within the first 1-2 Myr to account for the observed ratio of MS stars on the double MSs in young open clusters. Similarly, recent simulations on binary formation suggest that a significant fraction of early mergers (or even pre-MS mergers) occurs in B-type binaries, with estimates as high as ~30\% \citep{2020MNRAS.491.5158T}.

All stellar models are assumed to rotate at 55\% of their critical velocities at the ZAMS. This choice is motivated by the need to explain the major (red) MS stars in young star clusters \citep{2022NatAs...6..480W}.

The same physics used for the single star models is applied to our binary models.
Binary evolution begins at the ZAMS with circular orbits, and mass and angular momentum are transferred via Roche-lobe overflow \citep{1967ZA.....65..251K,2013ApJ...764..166D}. We account for the spin-up of accretors during mass transfer \citep{1981A&A...102...17P}, assuming that once an accretor reaches critical rotation, further accretion is halted. This results in accretion efficiency being dependent on the binary’s orbital period, with values below a few percent in long-period Case B binaries and reaching up to 60\% in short-period Case A binaries \citep{2022NatAs...6..480W,2022A&A...659A..98S}.

The excess material that cannot be accreted by the accretor is then assumed to be expelled from the stellar surface, driven by the combined radiation energy of both stars. If this radiation energy is insufficient to expel the material, we assume the two stars merge. Mergers are also assumed when mass is lost through the second Lagrangian point or when the mass transfer rate exceeds an ad-hoc upper limit of $\log \dot{M} > 10^{-1}\,M_\odot\,\mathrm{yr^{-1}}$, indicating unstable mass transfer. 

We use the method in \cite{2016MNRAS.457.2355S} to calculate the evolution of merger products from two MS stars, assuming that the merger products have an initial rotational velocity of 15\% of their critical value.
We do not model the evolution of merger products involving a post-MS star, as no well-established models currently exist. Previous studies suggest that such mergers are likely to ignite helium as blue supergiants (BSGs) or yellow supergiants (YSGs) \citep{1992PASP..104..717P,2014ApJ...796..121J} and may stay blue for most of their remaining lifetimes. We discuss this omission in Section\,\ref{sec:discuss_post_MS_merger}.

If none of the previously mentioned merger criteria are met, we follow the evolution of the binary models until core carbon exhaustion in both components. For numerical stability reasons, we limit the calculations to central helium exhaustion if the helium core mass exceeds $13\,M_\odot$. In some non-interaction binaries, the primary star encounters convergence issues during its asymptotic giant branch (AGB) evolution. For these cases, we terminate the evolution of the primary star at the end of helium exhaustion and allow the secondary star to continue evolving in isolation.

If the final core mass of the primary star surpasses the  Chandrasekhar limit at carbon/helium depletion, we assume a compact object formation, after which the companion is modeled as an isolated star. It is important to note that our simulations do not account for the impact of SN kicks. Our simulations represent the case in which binaries are disrupted due to the SN kick, with the secondary star remaining in the cluster as a single star. We discuss the impact of SN kicks in Section\,\ref{sec:number} and Appendix\,\ref{app_sec:SN_kicks}.

%===================================================================================================
\section{Results}\label{sec:Results}
%===================================================================================================
\subsection{Single-star models fail to explain observations}
\begin{figure}[htbp]
%FIG.1
\centering
\includegraphics[width=\linewidth]{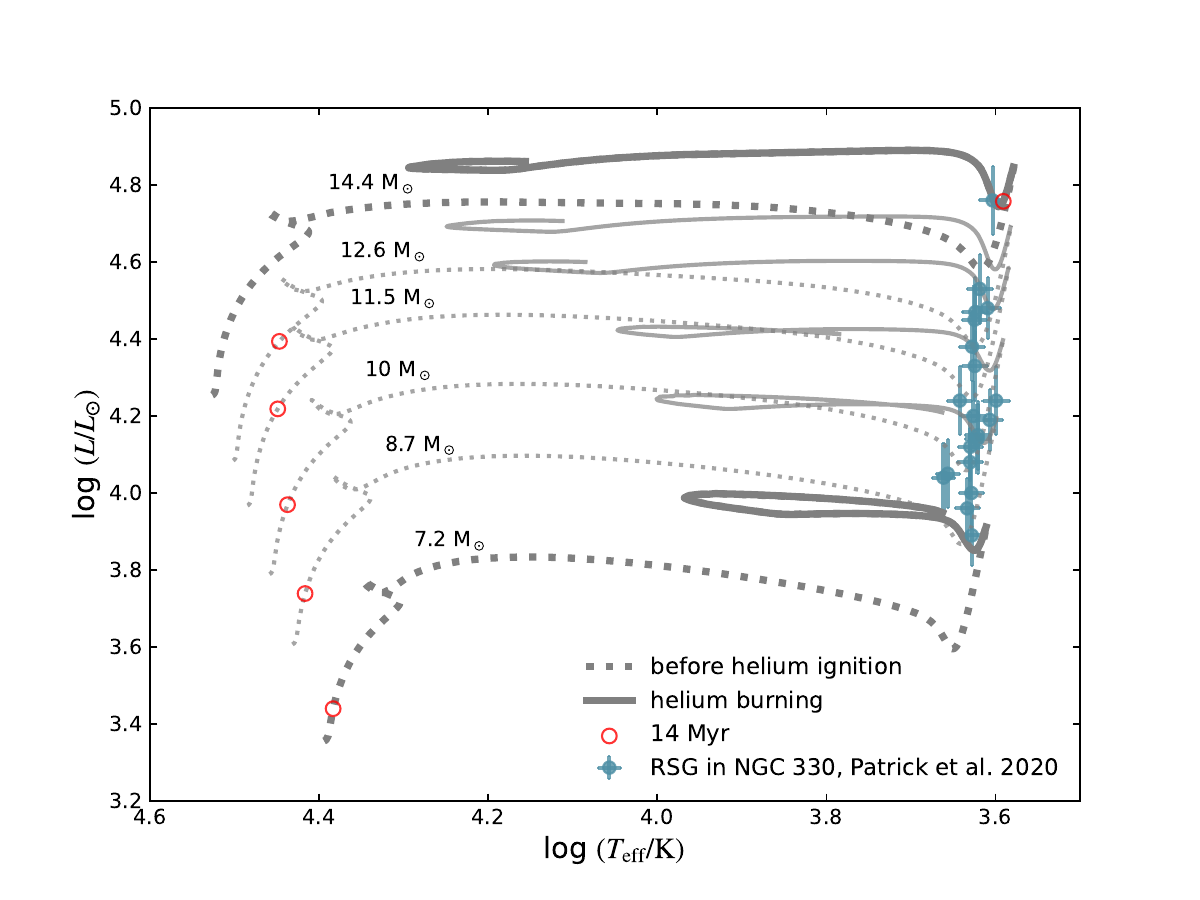}
\caption{Comparison of observed RSGs (cyan dots with error bars) in the SMC cluster NGC\,330 \citep{2020A&A...635A..29P} with the evolution of our non-rotating single-star models (grey lines) in the HRD. Initial stellar masses are indicated along the tracks. The lower and upper thick tracks represent the best fits for the least luminous and most luminous RSGs observed in NGC\,330, respectively. Solid grey lines denote the helium-burning stages of the stellar models, while dotted lines indicate the stages before helium ignition. Here we define the helium burning phase as the period during which the central helium mass fraction ranges from 0.98 and 0.001. The red circles indicate the positions of the stellar models at 14\,Myr for each mass. 
}
\label{fig:n330_single_track}
\end{figure}

In a star cluster comprised solely of coeval, non-rotating single stars, RSGs are expected to originate from stars with similar initial masses. Given the small luminosity variation of single stars during the helium-burning phase, the luminosity distribution of RSGs in a star cluster resulting from single-star evolution is anticipated to be narrow. However, recent spectroscopic observations of RSGs in young open clusters reveal a significantly broader luminosity range than predicted by single-star evolution \citep{2019A&A...624A.128B,2019MNRAS.486..266B}, based on Brott models \citep{2011A&A...530A.115B} and MIST models \citep{2016ApJ...823..102C}.

Using the SMC cluster NGC\,330 \citep{2020A&A...635A..29P} as an example, we compare the distribution of its RSGs with the evolutionary tracks from our MESA single-star models in the Hertzsprung-Russell diagram (HRD) in Fig.\,\ref{fig:n330_single_track}. It is evident that the observed luminosity spread of RSGs cannot be accounted for by one single-star track. Notably, the most luminous RSG requires a model nearly twice as massive as the least luminous one. Based on these models, the estimated ages of the least and most luminous RSGs are around 40\,Myr and 14\,Myr, respectively. This challenges our traditional understanding of star clusters that all stars are coeval. At 14\,Myr, stars below 14$\mso$ are still on their MS evolution. This indicates that the more massive, ``younger'' RSGs should have experienced rejuvenation to remain observable as a RSG at a similar age as the less massive star. Rejuvenation refers to a process in which a star appears younger than it actually is, primarily due to mechanisms that either mix fresh fuel into the core or simultaneously increase the star's total mass, thereby shortening its MS lifetime. The first scenario includes rotationally induced mixing \citep{Maeder2000,2009A&A...497..243D}, while the second includes binary mass transfer and binary mergers \citep{1977PASJ...29..249N,1983Ap&SS..96...37H,2016MNRAS.457.2355S}.

Our single-star models do not include the AGB phase. By analyzing our non-interaction binaries that successfully evolve through this phase, we found that
the probability of having an AGB star that is 0.5 dex more luminous than the least-luminous RSG in a star cluster is less than approximately 2\% (see Fig.\,\ref{appfig:AGB}). Therefore, AGB stars are unlikely to account for the observed luminosity spread of RSGs in young open clusters. Nonetheless, in rare cases, AGB stars can contaminate the red straggler population, as shown in our animation (e.g. see 26.86\,Myr).

The same result that the observed luminosity spread cannot be explained by single-star evolution is also found for RSGs in three LMC clusters, NGC\,2004, NGC\,2100 and NGC\,1818 (see Fig.\,\ref{appfig:3cluster_single_track}).

\subsection{Predictions from binary models and comparisons with observations}
\subsubsection{Distribution in the HRD}\label{sec:HRD}

\begin{figure}[htbp]
%FIG.1
\centering
\includegraphics[width=\linewidth]{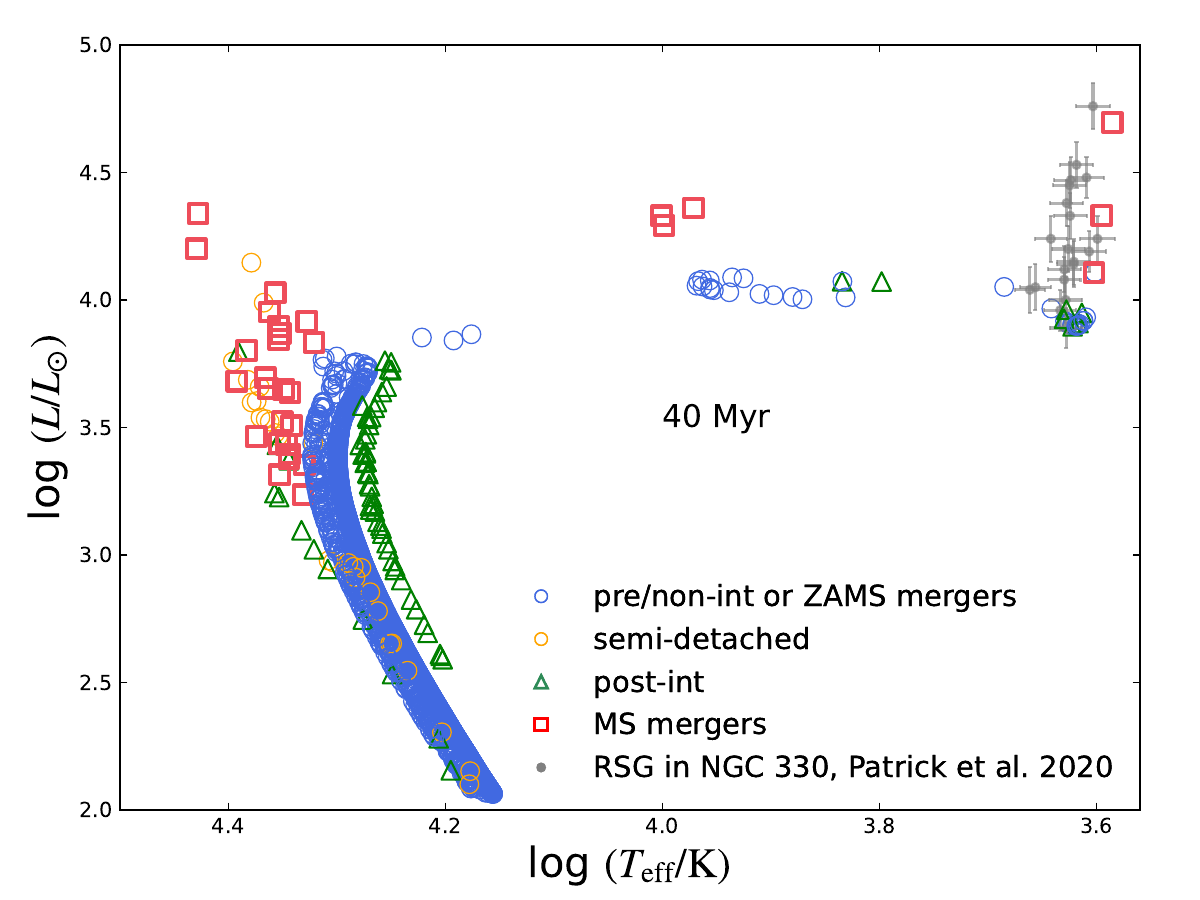}
\caption{Distribution of our detailed binary models in the Hertzsprung-Russell diagram at 40\,Myr. Each open marker indicates the visually brighter component of a binary system. Blue circles represent binaries that have not yet begun to interact or will never interact. 
Orange circles denote stars currently undergoing nuclear-timescale mass accretion. Green triangles mark stars that have experienced stable mass transfer. Red squares depict the products of mergers of two main-sequence stars. For comparison, we have superimposed the observed RSGs in NGC\,330, shown with grey dots with errorbars. This figure is a snapshot from an animation illustrating the distribution of our models from 2 to 100\,Myr. The animation is designed with a black background for enhanced visibility during presentation and has a duration of 27 seconds. The brown shaded area in the animation highlights the region where we identify RSGs. The full animation is available for download from the online journal.
}
\label{fig:n330_binary_HRD.pdf}
\end{figure}

In the following, we explore how binary evolution impacts the properties of RSGs in young open clusters. We show the distribution of our detailed binary models in the HRD at an age of 40\,Myr, alongside the observed RSGs in NGC\,330 \citep{2020A&A...635A..29P} in Fig.\,\ref{fig:n330_binary_HRD.pdf}. This age was selected because the luminosities of both the least and most luminous RSG models match the observations well, and it aligns with the age derived from the MS turn off of this cluster \citep{2022NatAs...6..480W}.

RSGs originating from pre/non-interaction binaries, as well as ZAMS mergers, evolve similarly to single stars. In this work, RSGs from pre-interaction binaries specifically refer to those that will undergo Case C mass transfer after helium exhaustion, which we assume to be unstable and lead to mergers. RSGs in pre-interaction binaries are primarily the primary stars in these binaries, while RSGs in non-interaction binaries can originate from either the primary or the secondary star. As anticipated, RSGs originating from these scenarios exhibit comparable luminosities, representing the least luminous subset of the RSG population.

In our simulations, RSGs formed from stable mass transfer exhibit similar luminosities to those from pre/non-interaction binaries. The reason is that the 
vast majority of these RSGs are accretors that have undergone Case B mass transfer (see the right panel of Fig.\,\ref{appfig:ex} for an evolutionary example), during which they accrete only tiny amounts of material. As a result, neither their total mass nor core mass increase significantly, leading to minimal rejuvenation.
On the other hand, accretors in Case A mass transfer (see the left panel of Fig.\,\ref{appfig:ex} for an evolutionary example) can accrete more material, leading to substantial increase in both their total and core masses, resulting in stronger rejuvenation. However, due to the presence of a composition gradient above the convective core, and given our assumptions on internal mixing, the accretor cannot undergo complete rejuvenation, where its core mass fully grows to match that of a genuinely single star with the same total mass. Instead, the accretor experiences incomplete rejuvenation, meaning it retains a smaller core compared to a genuinely single star of equivalent total mass. As shown in the left panel of Fig.\,\ref{appfig:ex}, such an accretor ignites helium and explodes as a BSG or YSG (see also \citealt{1995A&A...297..483B,2019A&A...621A..22F}).

Figure\,\ref{fig:n330_binary_HRD.pdf} demonstrates that, in our simulations, nearly all high-luminosity RSGs (i.e. red stragglers) are from MS mergers. MS merger products manifest as blue stragglers during their MS evolution, appearing younger and more massive than the MS turn-off stars. These blue stragglers will similarly appear younger and more massive than typical RSGs and appear as red stragglers when they evolve into the RSG phase. The more massive the MS merger product, the higher the luminosity it will have during its RSG phase.
It can be seen that our binary models closely reproduce the observed luminosity spread of RSGs in NGC\,330. This means that the observed luminosity of red stragglers can be explained by MS merger origin.

Figure\,\ref{fig:n330_binary_HRD.pdf} also shows that we predict a luminosity spread among BSGs, with MS merger products exhibiting higher luminosities. Notably, this spread has been observed in BSGs in young open clusters \citep{2018MNRAS.477.2640M}. A detailed comparison between the predicted and observed properties of BSGs is beyond the scope of this study.

We present an animation that illustrates the distribution of RSGs in the HRD up to 100\,Myr, with Figure\,\ref{fig:n330_binary_HRD.pdf} capturing a specific snapshot from this sequence.

\subsubsection{Luminosity of RSGs originating from MS mergers}\label{sec:RSG_MS_Merger}

 \begin{figure*}[htbp]
%FIG.1
\centering
\includegraphics[width=\linewidth]{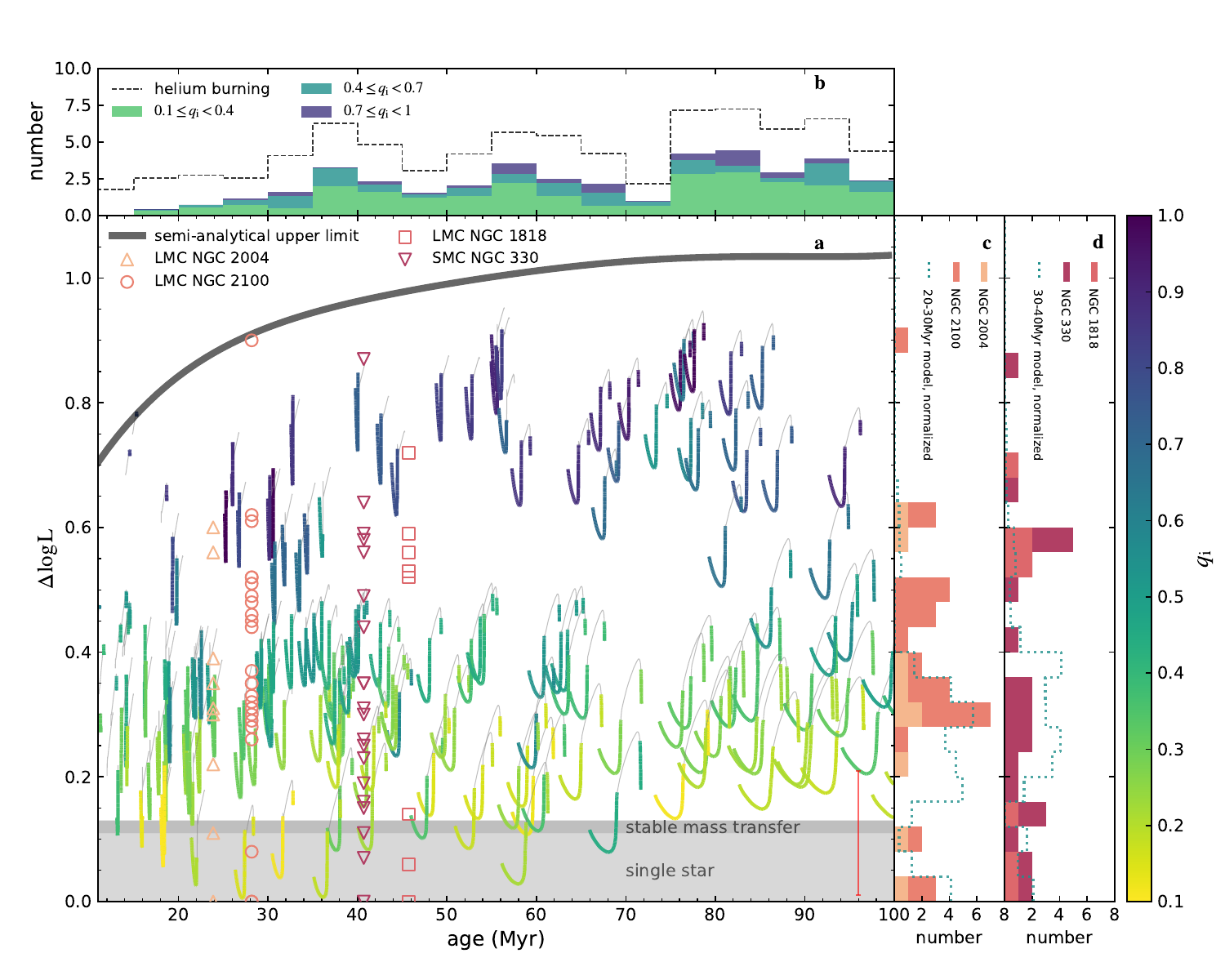}
\caption{Panel a: The luminosity extension $\Delta \log L=\log L -\log L_0$ for RSGs originating from MS mergers in our simulations as a function of age, where $L$ is the luminosity of RSGs and $L_0$ is the baseline luminosity (see Appendix\,\ref{app_sec:L0}). 
Each track represents the luminosity variation for one RSG, with color indicating the initial mass ratio $q_i$ (defined as the ratio of the secondary star's mass to the primary star's mass) of its binary progenitor (see the color bar on the right side of the figure). Thin grey lines along the tracks indicate periods when the corresponding MS merger model is in the BSG or YSG phase. The thick dark gray curve represents results from the semi-analytical approach where we assume mergers happen between two turn-off MS stars (see Appendix\,\ref{app_sec:semi_analytical}). Light and dark gray shade areas indicate the range of $\Delta \log L$ predicted by single-star evolution and stable mass transfer, respectively. These ranges are determined to encompass 95\% of the RSGs produced in these two scenarios. Open markers illustrate the luminosity difference for each observed RSG relative to the least luminous RSG in its cluster, with cluster ages determined by comparing this least luminous RSG with our single-star models (see Fig.\,\ref{appfig:RSG_L0}). The red error bar in the lower right corner of this panel represents the observational uncertainty ($\pm 0.1$\,dex). Panel b: The number distribution of RSGs in clusters with a total initial mass of $1.3\times 10^5\mso$ originating from MS mergers as a function of age. Models are categorized into three groups based on the initial mass ratios of the merger progenitors. The bars are stacked. The black dashed steps indicate the predicted number distribution for the helium-burning stars, i.e. including RSGs, BSGs and YSGs. Panel c: $\Delta \log L$ distribution for the observed RSGs in clusters NGC\,2004 (light orange bars) and NGC\, 2100 (orange bars). The bars are stacked. The green dashed steps depict the predicted $\Delta \log L$ distribution for our MS merger models between 20 and 30\,Myr, normalized to match the total number of observed RSGs in these two clusters. 
Panel d: Same as Panel c, but for observed RSGs in clusters NGC\,1818 (light red bars) and NGC\, 330 (dark red bars). And the green dashed steps represent the predicted $\Delta \log L$ distribution for our models between 40 and 50\,Myr.
}
\label{fig:delta_L_merger}
\end{figure*}

As shown in Figure\,\ref{fig:n330_binary_HRD.pdf}, our MS merger models successfully reproduce the observed luminosity extension of RSGs in NGC\,330.
In this section, we examine more generally the RSG luminosity extension that MS mergers can generate in young open clusters up to 100\,Myr and compare this with observational data. The results are shown in Figure\,\ref{fig:delta_L_merger}. 
Here $\Delta \log L=\log L-\log L_0$ represents the difference in logarithmic space between the luminosity of RSGs in our binary-merger models ($L$) and the reference luminosity ($L_0$) derived from single-star models at a given cluster age. The definition of $L_0$ is explained in detail in Appendix\,\ref{app_sec:L0}. 
Briefly, we use the weighted mean luminosity of single-star models during their RSG evolution as the baseline luminosity. 

Each track in Panel a represents the evolution of one RSG model. The changes in $\Delta \log L$ result from two combined effects: (1) the decrease of $\log L_0$ with age and (2) the luminosity evolution during the RSG phase. For reference, we also indicate the range of $\Delta \log L$ predicted by our models for RSGs originating from single-star evolution (light grey shaded area) and stable mass transfer (dark grey shaded area). To determine this, we analyze the results in Figs.\,\ref{fig:delta_L_accretion} and \ref{appfig:delta_L_primary} and calculate the range of $\Delta \log L$ encompassing 95\% of RSGs from these scenarios. Our analysis shows a range of -0.03 to 0.11 (a spread of 0.14 dex) for scenarios resembling single-star evolution, and -0.05 to 0.13 (a spread of 0.18 dex) for stable mass transfer. Accordingly, the boundaries of the light and dark grey shaded area are set to 0.11 and 0.13, respectively. It can be seen that MS mergers produce RSGs with a significantly broader luminosity distribution  compared to those produced by single-star evolution or stable mass transfer.

We notice that our simulations are limited to a single set of initial parameters and therefore cannot fully represent all star clusters. To remedy this, we provide a semi-analytical analysis to establish the upper limit for $\Delta \log L$ of RSGs originating from MS mergers (see Appendix\,\ref{app_sec:semi_analytical} for more detail). The idea is that the most luminous (and massive) red straggler in a cluster was previously the most luminous (and massive) blue straggler, which should have formed through the merger of two turn-off stars. The results are shown by the thick solid grey curve in Panel a of Fig.\,\ref{fig:delta_L_merger}. We predict an upper limit of 0.8 to 1.0\,dex for the luminosity extension of RSGs in clusters aged between 15 and 100\,Myr. RSGs originating from MS mergers are not expected to exceed this luminosity limit.

In Panel a of Fig.\,\ref{fig:delta_L_merger}, we overplot observational data from four young open clusters, with $\Delta \log L$ representing the luminosity difference between each observed RSG and the least luminous RSG within the same cluster in the logarithmic space. The cluster ages are determined by comparing the luminosity of the least luminous RSG in each cluster with the relationship between $L_0$ and cluster age predicted by single-star models (see the orange lines in Fig.\,\ref{appfig:RSG_L0}). Since metallicity has a negligible impact on the relationship between $L_0$ and cluster age, it is reasonable to compare observations from both LMC and SMC with our SMC binary models. It can be seen that the luminosities of all observed RSGs fall within the predicted range, indicating that the observed high-luminosity red stragglers can be effectively explained by the MS merger scenario.

Panel b shows the number distribution of RSGs originating from MS mergers, as predicted by our simulations, grouped by the initial mass ratios of the merger progenitors.
The number in each age bin is calculated as 
\begin{equation}\label{eq:1}
N = \sum\limits_{\rm i} \frac{t_{\rm i}}{t_{\rm bin}},
\end{equation}
where $i$ represents the index of each RSG, $t_{\rm i}$ denotes the time the $i$-th RSG spends in this age bin, and $t_{\rm bin}=5$\,Myr is the width of the age bin.
It can be seen that the predicted number of RSGs does not vary significantly with age. On one hand, the influence of the IMF would suggest a higher number of RSGs in older clusters. On the other hand, the parameter space for MS mergers decreases with lower primary masses (see Fig.\,\ref{appfig:outcome}), and the mass range of RSGs within a constant age bin also narrows over time.

In Panels c and d, we present the $\Delta \log L$ distribution for the observed RSGs, divided into two age groups. 
We also show the predicted $\Delta \log L$ distributions of RSGs originating from MS mergers for the specific age ranges indicated in the figure legend. These distributions are normalized such that the total number of predicted RSGs matches the number of observed RSGs in each panel. It is important to note that we do not intend to directly compare our model predictions with the observations here, as the predictions include only MS merger products. If contributions from scenarios resembling single-star evolution were included, a significant peak at $\Delta \log L\lesssim 0.1$ (see Fig.\,\ref{appfig:delta_L_primary}) would be expected. A comparison that accounts for RSGs from other evolutionary scenarios is provided in Fig.\,\ref{fig:RSG_number}. Here, we focus solely on the $\Delta \log L$ distribution of RSGs from MS mergers. 

Our predictions align with observational data in that the majority of RSGs exhibit relatively low luminosities ($\Delta \log L \lesssim 0.4$ in our predictions, compared to $\Delta \log L \lesssim 0.6$ in observations). In our simulations, lower-mass-ratio binaries are more likely to merge than higher-mass-ratio binaries, leading to a prediction of more lower-luminosity RSGs than higher-luminosity RSGs originating from the MS merger scenario. The discrepancy between the predicted peak at $0.2 \lesssim \Delta \log L \lesssim 0.4$ and the significant fraction of observed RSGs with $0.4 \lesssim \Delta \log L \lesssim 0.6$ may be explained by observational errors ($\pm$0.1 dex), photometric variability and the accuracy of the measurement of the least luminous RSG in these clusters. We will discuss this in detail in Section\,\ref{sec:intermediate_L_RSG}.

\subsubsection{Luminosity of RSGs originating from stable mass transfer}\label{sec:L_stable_MT}

 \begin{figure}[htbp]
%FIG.1
\centering
\includegraphics[width=1.1\linewidth]{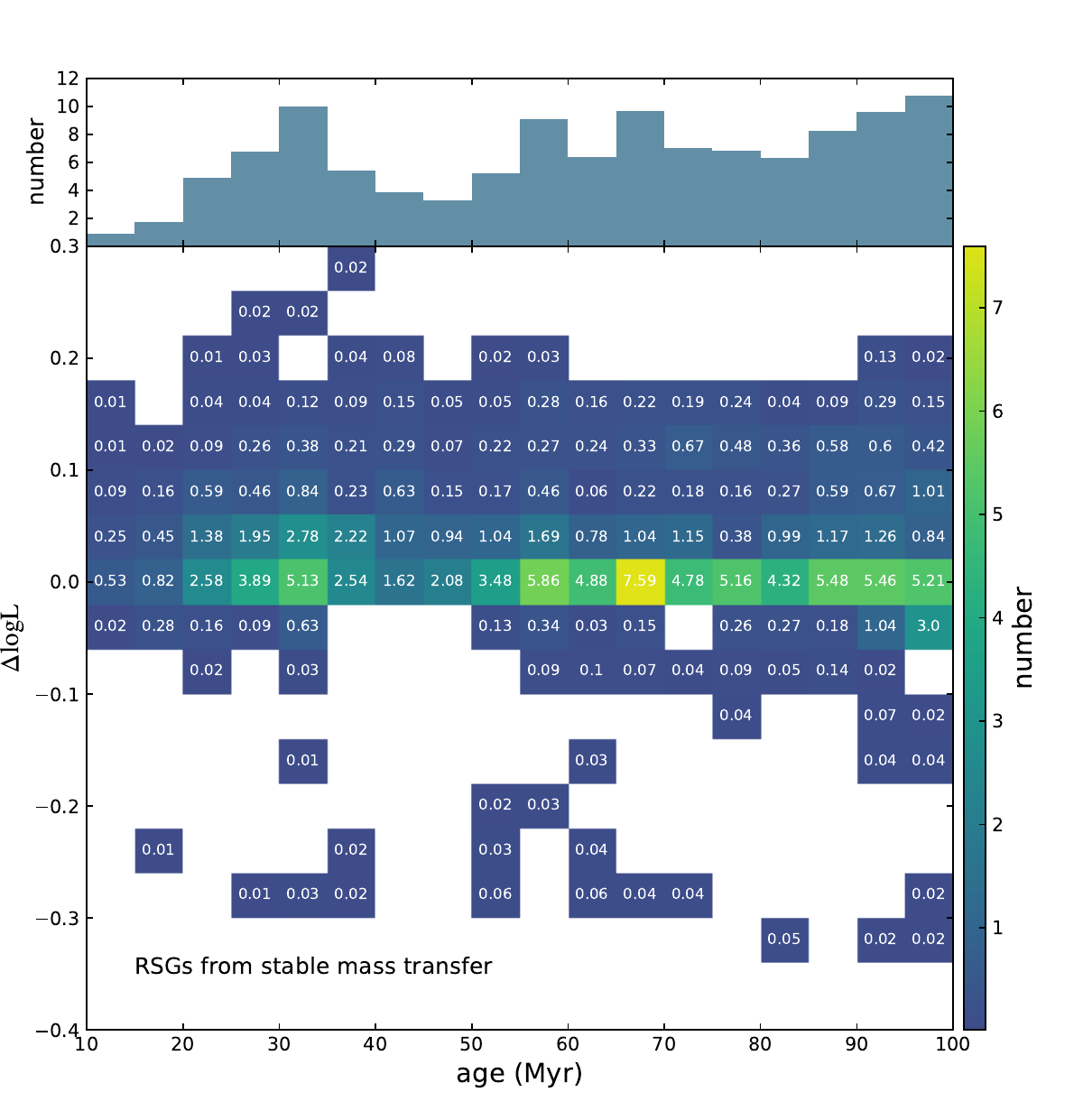}
\caption{Predicted luminosity extension $\Delta \log L=\log L -\log L_0$ for RSGs originating from stable mass transfer as a function of age. The color scale indicates the predicted number of RSGs in clusters with a total initial mass of $1.3\times 10^5\mso$ produced through stable mass transfer. Exact value in each pixel is labeled. 1D projections of the distribution are displayed in the top panel. 
}
\label{fig:delta_L_accretion}
\end{figure}

In Fig.\ref{fig:delta_L_accretion} we display the predicted $\Delta \log L$ for RSGs originating from stable mass transfer across different cluster ages. Unlike Fig.\,\ref{fig:delta_L_merger}, here we use the heat plot and display the predicted number of RSGs in each pixel, because RSGs originating from stable mass transfer have a much narrower $\Delta \log L$ distribution and a larger numbers, such that all tracks like those in Fig.\,\ref{fig:delta_L_merger} will accumulate at $\Delta \log L \sim 0$. The number in each pixel is calculated using Eq.\ref{eq:1}. It is important to note that in our standard simulations, we assume that the secondary star evolves in isolation and remain in the cluster after the primary star evolves to a compact object.

Consistent with Fig.\,\ref{fig:n330_binary_HRD.pdf}, Fig.\,\ref{fig:delta_L_accretion} demonstrates that the luminosity extension of the RSGs originating from stable mass transfer is small due to low mass transfer efficiency, with 95\% having $\Delta \log L$ between -0.05 and 0.13 (0.18 dex spread). RSGs with $\Delta \log L \lesssim -0.1$ are those before helium ignition and are unlikely to have significant observational consequences (see also Fig.\,\ref{appfig:delta_L_primary}).

\subsubsection{Number of RSGs and fraction of red stragglers}\label{sec:number}
\begin{figure*}[htbp]
%FIG.1
\centering
\includegraphics[width=\linewidth]{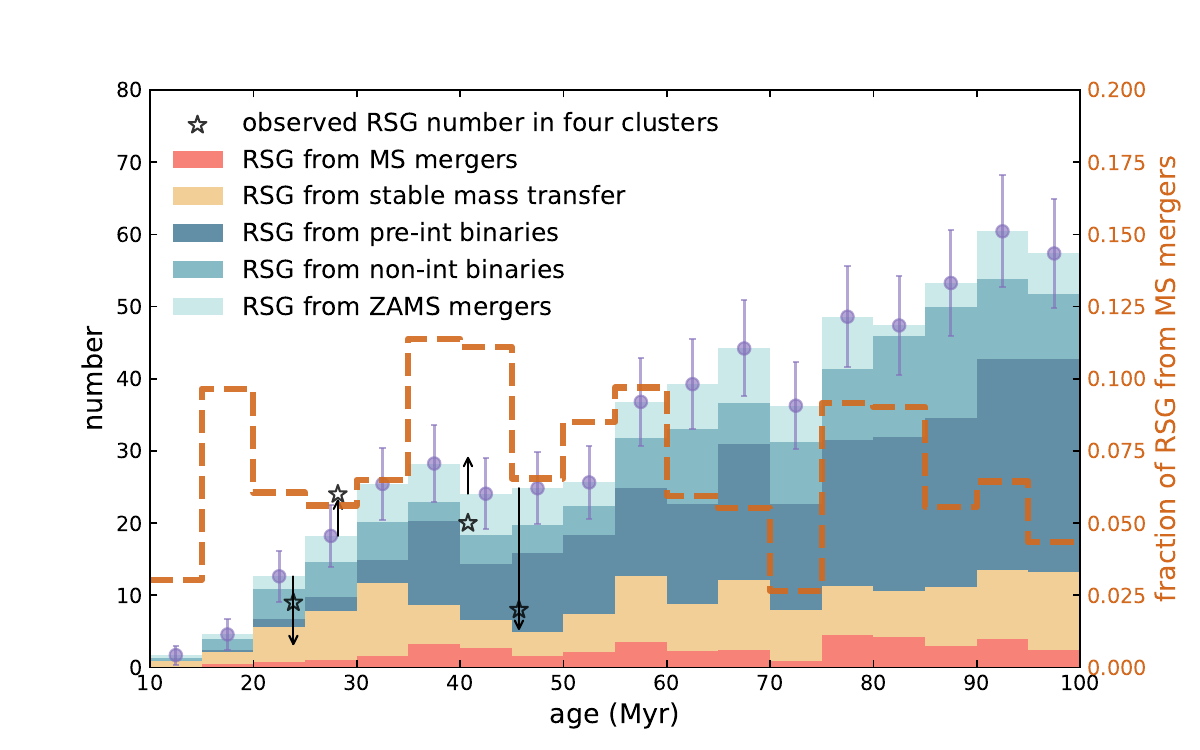}
\caption{Number of RSGs in clusters with an initial total mass of $1.3\times 10^5 \mso$ from different evolutionary scenarios predicted by our detailed binary models. Orange-red bars represent RSGs originating from MS mergers. Yellow bars depict RSGs resulting from stable mass transfer without taking into account supernova kicks. Dark blue and medium blue bars represent RSGs that have not interacted with their companions or will never interact, respectively. Light blue bars correspond to RSGs from ZAMS mergers. 
The bars are stacked to provide a cumulative representation. Purple dots with errorbars indicate the standard deviation of the total number of RSGs predicted in our simulations. Open pentagrams show the observed number of RSGs in clusters NGC\,2004, NGC\,2100, NGC\,330, NGC\,1818, displayed from left to right. The endpoints of the black arrows show the predicted number of RSGs for each cluster, adjusted according to their total masses (see the text for details). The brown dashed steps illustrate the fraction of RSGs from MS mergers relative to the total number of RSGs in our simulation (i.e., the value predicted by the orange-red bars divided by the sum of all five populations). The corresponding values are shown on the right y-axis.
}
\label{fig:RSG_number}
\end{figure*}

In Figure.\,\ref{fig:RSG_number} we show the predicted number of RSGs in clusters with initial stellar masses of $1.3\times 10^5\mso$ arising from different evolutionary scenarios. The number is calculated using the same method as that in Panel b of Fig.\,\ref{fig:delta_L_merger}. 

It can be seen that the number of RSGs from pre/non-interaction binaries increases with cluster age, due to the IMF and the increasing parameter space for non/pre-interaction systems (see Fig.\,\ref{appfig:outcome}). In contrast, the number of RSGs formed from stable mass transfer and MS mergers remains relatively constant with cluster age. As previously explained, the effect of the IMF is balanced by the shrinking parameter space for these evolutionary pathways. 

To compare our predicted numbers with observational data, we scale the predictions by the ratio of the observed cluster mass to the simulated mass ($1.3\times10^5 \mso$). The scaled results are represented by the ends of the black arrows in Fig.\,\ref{fig:RSG_number}. For the observational cluster masses, we use the following values: $1.58\times 10^5 \mso$ for NGC\,330 \citep{2020A&A...635A..29P}, $2.3\times 10^4 \mso$ for NGC\,2004 \citep{2005ApJS..161..304M,2015A&A...575A..62N}, $1.5\times 10^5 \mso$ for NGC\,2100 \citep{2016MNRAS.458.3968P} and $2.57\times 10^4 \mso$ for NGC\,1818 \citep{2005ApJS..161..304M,2018MNRAS.477.2640M}. 
We remind readers that the scaled results are influenced by uncertainties in the estimated cluster masses, which themselves are highly uncertain. For instance, \cite{2016MNRAS.458.3968P} provided an upper limit on the dynamical mass of NGC\,2100, which is six times larger than the photometric mass reported in \cite{2005ApJS..161..304M}. Where available, for consistency we adopt dynamical masses for the clusters, which avoids assuming a mass-to-light ratio to determine the total mass of the cluster. Despite the sensitivity of the predicted number of RSGs to various factors (see Section\,\ref{sec:uncertain_phy}) our predictions align reasonably well with the observed numbers.

However, it is important to note that the actual number of RSGs originating from non-interaction binaries, ZAMS mergers and stable mass transfer could differ significantly from our simulation predictions. As explained in Section\,\ref{sec:method_binary}, the fractions of non-interaction binaries and ZAMS/pre-MS mergers in young open clusters are highly uncertain. Nevertheless, the evolution of stars in these two populations is identical to that of single stars. Consequently, the uncertainty in these populations can be effectively modeled by varying the assumed binary fraction in star clusters (see Appendix\,\ref{app_sec:fb}). 

The actual number of RSGs originating from stable mass transfer in young star clusters should be lower than shown in Fig.\,\ref{fig:RSG_number}, once the effects of SN kicks and reverse mass transfer are considered. In Appendix\,\ref{app_sec:SN_kicks} we demonstrate that SN kicks significantly affect stars in clusters aged approximately 15 to 50\,Myr. For binaries where the primary star undergoes a SN explosion, over 90\% are disrupted due to SN kicks (see also \citealt{2024ApJ...975L..20W}). The typical velocity imparted to the secondary star is on the order of tens of $\kms$, which may exceed the typical escape velocity (approximately $12-15 \kms$) of young open clusters \citep{2014ApJ...797...35G,2010ARA&A..48..431P}.
Consequently, the majority of these disrupted binaries are expected to produce runaway RSGs. However, there are currently no examples of RSG runaway stars identified in the MCs. Examining the cluster population of NGC\,330 as an example, between one and five RSGs could be considered candidate runaway stars based on their projected separation from the cluster centre \citep[see the distinction between the cluster and field populations in] []{2021A&A...647A.135N}. However, their Gaia~DR3 proper motions do not clearly corroborate this. 
Binaries that either do not experience SN explosions, i.e. those form black holes (BHs) or white dwarfs (WDs), or remain bound after the SN event are likely to undergo reverse mass transfer from the secondary star to the compact object, resulting in tripped helium stars before they reach the RSG phase. Accounting for these two effects, we find that, on average, 80\% to 90\% of the RSGs predicted in our standard simulations may not persist as RSGs in young open clusters aged between 20 and 50\,Myr (see Fig.\,\ref{appfig:SN_kicks}).

A more helpful comparison is examining the fraction of red stragglers, which are likely predominantly formed through MS mergers. The brown dashed line in Fig.\,\ref{fig:RSG_number} shows that the fraction of RSGs originating from MS mergers relative to those from all the considered scenarios remains below approximately 12\% across different ages, which is significantly lower than the observed value of approximately 63\% to 88\% when identifying RSGs with $\Delta \log L \ge 0.2$ as red stragglers.

%===================================================================================================
\section{Discussions}\label{sec:discussion}
%===================================================================================================

\begin{table*}[ht] % Force the table to appear exactly here
\centering
\caption{Summary of properties of RSGs originating from various evolutionary scenarios, as predicted by our single and binary models.}
\label{tab:1}
\begin{tabular}{p{3cm}p{6cm}p{5cm}p{3cm}}
\toprule
Pathway & Produce RSGs or BSGs? & RSG luminosity extension  & Remain in clusters? \\
\midrule
single star evolution  & can be both & $\sim$ 0.14 dex    & yes \\
pre/non-interaction binaries  & can be both & $\sim$ 0.14 dex      & yes \\
pre-MS/ZAMS mergers & can be both     & $\sim$ 0.14 dex   & yes \\
MS mergers       & can be both       & up to $\sim$ 1 dex       & yes \\
post-MS mergers  & most likely BSGs, hard to produce RSGs & likely resemble single-star evolution   & yes \\
Case B mass transfer & most likely runaway RSGs or blue stripped stars & $\sim$ 0.18 dex    & mostly no\\
Case A mass transfer  & BSGs or blue stripped stars, unlikely to produce RSGs & cannot exceed that of MS mergers  & - \\
\bottomrule
\end{tabular}
%\footnotesize{1. Frequency is defined as the ratio between the number of blue stragglers and all stars in a cluster.}
\end{table*}
%\linenumbers

\subsection{Red straggler fraction}\label{sec:red_straggler_frac}
We summarize the properties of RSGs evolving from different scenarios and their properties in Table.\,\ref{tab:1}. Our simulations suggest that MS mergers are likely the primary contributors to the population of red stragglers in star clusters.
Figure\,\ref{fig:RSG_number} shows that our standard simulations predict approximately 90\% of RSGs to have luminosities consistent with single-star evolution (with a spread of approximately 0.14 dex), while the remaining 10\% are predicted to have higher luminosities, contributing to the red straggler population. However, observations reveal that a much larger fraction of RSGs (63\% to 88\%) have luminosities at least 0.2 dex higher than the least luminous RSGs in young young open clusters. Excluding RSGs originating from non-interaction binaries, ZAMS mergers and stable mass transfer can increase the predicted red straggler fraction to 20-40\% between 20 to 50\,Myr. Nonetheless, it is unlikely that all these populations are absent, particularly given that the ZAMS/pre-MS merger fraction is expected to be higher than our simulations suggest \citep{2022NatAs...6..480W}. Additionally, this upper limit of 40\% is achieved at 20\,Myr, a point at which we already underpredict the total number of RSGs observed in NGC\,2004. Therefore, further removing populations from our standard simulations is not anticipated. Assuming a power-law initial orbital period distribution skewed toward short periods with $f(\log P)\propto \log P^{\pi}$ with $\pi = -0.55$ \citep{2012Sci...337..444S} could further increase the fraction of MS mergers (i.e., red stragglers) by a factor of about two. 

Moreover, triple star evolution has been proposed as a mechanism that can trigger mergers of inner binaries through the von Zeipel-Lidov-Kozai (ZLK) cycles \citep{1910AN....183..345V,1962P&SS....9..719L,1962AJ.....67..591K}, which could also increase the red straggler fraction. However, the fraction of triple systems and their contributions in star clusters remain poorly understood. Interestingly, \cite{2024arXiv241218554P} found that in a significant fraction (5 out of 16) of binaries consisting of a RSG and a MS B-type star, the RSG appears significantly younger than its MS companion. Such systems can be explained by triple star evolution, where the inner binary merges to form a red straggler that appears younger than its MS companion.

From the observational side, in NGC\,2004, NGC\,2100 and NGC\,330, there seems to be a boundary at $\Delta \log L=0.4$, above and below which the observed RSGs are grouped. Stars above this boundary are likely red stragglers originating from binary mergers. Using this threshold to identify red stragglers results in red straggler fractions of 23\% to 42\% in these three clusters, which align with our discussions above. RSGs with luminosities below this threshold are likely the result of scenarios resembling single-star evolution. However, single-star evolution alone can account for only a luminosity extension of approximately 0.14 dex. To explain these intermediate-luminosity RSGs, an additional luminosity extension of approximately 0.26 dex from non-binary effects is required.

\subsection{Explanations for intermediate-luminous RSGs}\label{sec:intermediate_L_RSG}

\subsubsection{Observational uncertainties}
The observational uncertainty of 0.2 dex ($\pm$ 0.1 dex) used in this paper is derived by comparing luminosities determined from SED fitting with the technique of \cite{2013ApJ...767....3D} (see for example the comparison in \citealt{2024arXiv241218554P}). This value is smaller than the uncertainty of approximately 0.34 dex ($\pm$ 0.17 dex) obtained using the K-band magnitude or luminosity calibration presented in \cite{2013ApJ...767....3D}. However, it may still represent an overestimate (L. Patrick in private communication). 

Studies have shown that variability is common among RSGs, which can influence the accuracy of their luminosity measurements. In the V-band, variability can reach up to 4 magnitudes, equivalent to 1.6 dex in luminosity \citep{2006MNRAS.372.1721K,2020ATel13512....1G}. This large variability is because that the spectral appearance of RSGs in the visual bands are dominated by large TiO molecular features, which are well documented to be significantly variable. 
In contrast, variability in the near-infrared bands is believed to be substantially smaller \citep{2018ApJ...859...73S,2018A&A...616A.175Y}. A typical 
variability of 0.08\,mag is observed in the K-band (from which the RSG luminosities used in this study are derived), with extreme cases reaching up to 0.25\,mag \citep{1983ApJ...272...99W,2009ApJ...703..420M}. The extreme amplitude of this 0.25\,mag corresponds to a change in luminosity of 0.1 dex, which is well within the uncertainties on the luminosities. However, studies on the variability of RSGs in the near-infrared are still rare. Further studies are crucial for understanding how this variability contributes to the luminosity spread of RSGs in young open clusters.

\subsubsection{Rotational mixing}
Rotational mixing is proposed to be an important mechanism to drive stellar rejuvenation \citep[see][for a review]{Maeder2000}. In our models, rotational mixing is inefficient, such that the MS lifetime of moderately rotating stars differs only slightly (by less than a few percent) from that of non-rotating stars. In contrast, in models with efficient mixing, such as the Geneva models \citep{2012A&A...537A.146E}, the MS lifetime can be extended by up to about 25\%.
This implies that fast-rotating stars ignite helium later than their non-rotating counterparts, and may have higher luminosities during their RSG phase. In Appendix\,\ref{app_sec:rotation} we explore the effects of stellar rotation by analyzing Geneva models. We find that while efficient rotational mixing can lead to enhanced rejuvenation and longer MS lifetimes, fast-rotating stars generally remain blue during their helium-burning phases and are thus unlikely to contribute significantly to the RSG population. It is worth noting that in Geneva models with masses above $10\mso$, even non-rotating stars remain mostly blue. Whether a star manifests as a BSG or RSG is highly sensitive to internal mixing processes, which remain poorly constrained. Future theoretical studies, particularly on the combined effects of various internal mixing processes, are 
essential for a comprehensive understanding of the role of rotational mixing plays in the luminosity spread of RSGs in young open clusters. 

\subsubsection{Real age difference}
Real age difference can also cause a luminosity extension of RSGs in young open clusters. It is important to note that unlike old and massive globular clusters, which show evidence of multiple stellar generations \citep{2012A&ARv..20...50G,2018ARA&A..56...83B}, current studies of young open clusters argue against significant real age differences among their stars (see Appendix\,\ref{app_sec:age_diff} for detail). 
However, small age differences may exist in young open clusters (less than approximately 5\,Myr, see Fig.\,\ref{appfig:CMD_delta_age_diff}) . \cite{2022NatAs...6..480W} proposed that approximately 10-30\% of binaries may merge due to orbit decay within the first $\sim$2\,Myr after their formation. 
Such early mergers experience rejuvenation, making them appear up to 2 Myr younger than other stars. While this is not a true age difference, it effectively mimics one because it occurs early in the cluster evolution and the fraction of such early mergers is high. 
In Fig.\,\ref{appfig:deltaL_age_diff}, we illustrate that a real age difference of 2\,Myr can lead to a luminosity extension ranging from 0.03 to 0.07 dex within the age range of the observed clusters. A real age difference of 5 Myr can result in a significantly larger luminosity extension, spanning from 0.1 to 0.18 dex within the same age range. When combined with contributions from single-star evolution ($\sim 0.14$ dex) and observational uncertainties (0.2 dex), the total luminosity extension amounts to approximately 0.4 to 0.5\,dex. This is sufficient to explain the observed population of low- to intermediate-luminosity RSGs in young open clusters. 

\subsubsection{Measurement accuracy of the least-luminous RSG}
The luminosity extension of the observed RSGs is determined by the least-luminous RSG in these clusters. In NGC\,2004, NGC\,2100 and NGC\,1818, the two to three least-luminous RSGs are noticeably less luminous than the luminosity where a significant fraction of RSGs accumulate ($\Delta \log L$ ranges from 0.25 to 0.4 in NGC\,2004 and NGC\,2100, and from 0.5 to 0.6 for NGC\,1818). Excluding these outliers would significantly improve the agreement between predictions and observations. Therefore future studies to confirm whether these two to three least luminous RSGs are genuine cluster members and to accurately measure their luminosities are crucial.

In summary, we argue that the observed low- to intermediate-luminosity RSGs (with extensions up to approximately 0.4-0.5 dex) are likely the result of processes resembling single-star evolution. In contrast, RSGs with higher luminosities are more likely red stragglers originating from MS mergers. It is important to note that the uncertainties mentioned above also apply to the interpretation of red stragglers, as our simulations predict their luminosity peak at $\Delta \log L$ between 0.2 and 0.4.

\subsection{RSGs as indicators of cluster age}\label{sec:discussion_age}

Estimating the age of young open clusters is a critical yet challenging task, particularly due to the presence of an extended MS turn-off, which complicates the traditional approach of using the MS turn-off as an age indicator. \cite{2019MNRAS.486..266B} and \cite{2019A&A...624A.128B} proposed that the least-luminous RSG in a cluster serves as a reliable age indicator, as it likely represents the oldest RSG in the cluster and is expected to be a product of single-star evolution. \cite{2020MNRAS.495L.102E} examined BPASS single and binary models and suggested that the mean luminosity of the entire RSG population is a better age indicator. They argued that while binary evolution can produce a larger luminosity spread among RSGs, the mean luminosities derived from single and binary scenarios remain nearly identical. Consequently, they advocated for using the entire RSG population rather than just the least-luminous RSG.

We disagree with the conclusion of \cite{2020MNRAS.495L.102E}, as our simulations indicate that the high-luminosity RSG population is predominantly composed of binary merger products. \cite{2020MNRAS.495L.102E} examined the luminosity ranges produced by single and binary evolution but did not account for the relative time a star spends at each luminosity. Our population synthesis simulations based on detailed binary models demonstrate that MS mergers are the primary contributors to the luminosity spread of RSGs. 

To determine cluster age, it is essential to focus on RSGs originating from scenarios resembling single-star evolution. Stellar evolutionary models predict a small luminosity spread of approximately 0.14\,dex for these scenarios. However, when comparing these predictions with observed distributions of RSGs, we argue that additional luminosity extension, potentially caused by observational errors and real age differences, must be considered. As explained in Section\,\ref{sec:intermediate_L_RSG}, low- to intermediate-luminosity RSGs (with extensions up to approximately 0.4-0.5\,dex) are likely the result of processes consistent with single-star evolution. Therefore, we propose that the mean luminosity of this population serves as a more reliable indicator of cluster age.

\subsection{The impact of uncertain physics}\label{sec:uncertain_phy}

\subsubsection{single star evolution}\label{sec:uncertain_single}

The evolution of massive stars into RSGs and the duration of this phase are sensitive to stellar internal mixing and wind mass loss during the RSG phase, both of which are still not well understood \citep{2019A&A...625A.132S,2010ApJ...717L..62Y,2014ARA&A..52..487S,2024arXiv241007335Z}. 

In this work, we adopt efficient semi-convective mixing with $\alpha_\mathrm{SC}=10$. \cite{2019A&A...625A.132S} shows that for $\alpha_\mathrm{SC}\geq 1$, the evolutionary outcomes remain similar. We use a mass-dependent overshooting parameter that ranges from 0 to 0.33. As shown in Fig.\,\ref{appfig:RSG_single}, our SMC models with masses between 5 and 16.6$\mso$ evolve into RSGs and subsequently undergo a blue loop. Star models above 19$\mso$ ignite helium as BSGs, with only some evolving to the red side during helium-burning stage. For the LMC models, stars with masses below approximately 15$\mso$ do not experience a blue loop. Varying these internal mixing parameters can influence the occurrence of a blue loop \citep{2019A&A...625A.132S}.
While this uncertainty can affect the predicted number of RSGs, it does not influence the luminosity of RSGs. Therefore, our conclusion that red stragglers may result from MS mergers remains unaffected.

Mass loss of RSGs also remains poorly constrained, particularly at low metallicities. This uncertainty does not significantly influence the luminosity of RSGs, but it has a substantial impact on their lifetimes. Consequently, the predicted luminosity extension of RSGs is unaffected by this uncertainty, but the predicted number of RSGs can vary.
 \cite{2024arXiv241007335Z} explored the impact of varying mass-loss rates on the formation and evolution of RSGs in the SMC. They found that higher mass-loss rates, such as those in \cite{2021A&A...646A.180K} and \cite{2023A&A...676A..84Y}, lead to earlier envelope stripping and shorter RSG lifetimes. In contrast, lower mass-loss rates, such as those in \cite{2020MNRAS.492.5994B}, can extend the RSG phase. 
\cite{2023A&A...676A..84Y} derived an empirical mass-loss prescription for RSGs based on a large sample of stars in the SMC, which is slightly higher than the rates used in our models. Adopting this empirical prescription would likely reduce the predicted number of RSGs.

In summary, the predicted luminosity spread of RSGs, a key factor in constraining their origins, remains unaffected by uncertainties in single-star evolution. However, the total number of RSGs is sensitive to these uncertainties. Further comprehensive studies on comparing the observed and predicted ratio of RSGs to BSGs/YSGs in young open clusters may offer valuable constraints on uncertain mixing processes and mass loss prescriptions.

\subsubsection{mass transfer stability and efficiency}\label{sec:discussion_MT}
Mass transfer stability and efficiency remain the least understood aspects of binary evolution. Previous studies \citep{2021A&A...653A.144H,2022NatAs...6..480W} have pointed out that to explain the high number of Be X-ray binaries in the SMC \citep{2016A&A...586A..81H} and the large fraction of Be stars in young open clusters \citep{2018MNRAS.477.2640M}, a greater fraction of binaries must undergo stable mass transfer than is currently predicted by our models. In addition, a higher mass transfer efficiency than predicted by our models has also been suggested as essential for explaining the observed properties of Be X-ray binaries \citep{2014ApJ...796...37S,2020MNRAS.498.4705V,2024ApJ...971..133R}.

However, whether a higher mass transfer efficiency contributes to the population of red stragglers remains uncertain. On the one hand, whether the rejuvenated star evolves to a RSG or BSG depends on the efficiency of internal mixing processes. If the rejuvenation is complete, the star's subsequent evolution would resemble that of a single star, and may evolve into a RSG (this statement may apply to all processes capable of inducing rejuvenation, including stellar rotation). However, incomplete rejuvenation might result in the star evolving into a BSG rather than a RSG. On the other hand, as previously mentioned, after considering the effects of SN kicks and reverse mass transfer, the stable mass transfer scenario is not expected to significantly contribute to the RSG population in clusters younger than about 50\,Myr.

\subsubsection{post-MS mergers}\label{sec:discuss_post_MS_merger}
In our simulations, we do not track the evolution of the mergers involving a post-MS star, due to the lack of well-studied, detailed models. These mergers typically have a helium core from the post-MS star and an envelope that mixes the post-MS star's envelope with the entire MS star. Therefore, such merger products have a lower core-to-envelope mass ratio compared to single stars of equivalent total mass. Previous studies using simplified models suggest that these merger products are likely to ignite helium as BSG or YSG stars, and may remain in these phases without evolving into RSGs \citep{1995A&A...297..483B,2014ApJ...796..121J,2024ApJ...967L..39B,2024A&A...686A..45S}.

Excluding post-MS mergers from our analysis does not impact our predictions regarding the maximum luminosity of red stragglers in young open clusters, for several reasons. Firstly, in a cluster, the mass of RSGs from post-MS mergers cannot exceed that of mergers between two MS turn-off stars, as they experience stronger rejuvenation. 
Second, although the initial luminosity of post-MS merger products depends on their total mass, their luminosity is expected to decrease as they evolve towards the RSG phase, where luminosity is primarily determined by core mass \citep{2024A&A...686A..45S}. Post-MS mergers generally have smaller core masses than MS mergers due to weaker rejuvenation. Finally, \cite{2014ApJ...796..121J} suggested that merger products with a lower core-to-envelope ratio remain blue for longer, meaning that more massive post-MS mergers from higher mass-ratio binaries, which could produce more luminous RSGs, are less likely to evolve into RSGs at all. In summary, excluding post-MS mergers does not affect the predicted luminosity range of RSGs formed through binary evolution, nor does it alter our conclusion that MS mergers are responsible for the most luminous RSGs in young open clusters.

However, excluding post-MS mergers may impact the predicted number of RSGs. \cite{2024A&A...686A..45S} show that post-MS mergers can evolve redwards, depending on the mass the primary star accreted and the detailed structure of the merger envelope. 
It is important to note that a significant amount of our models (around 50\% depending on stellar masses) merge in the post-MS phase (see Fig.\,\ref{appfig:outcome}). However, as previously mentioned, if these mergers do evolve into RSGs, they may contribute to low- to intermediate-luminosity RSGs. To accurately assess the impact of post-MS mergers on the population of RSGs in young open clusters, it is crucial to develop detailed models for these mergers. Additionally, a better understanding of the stability of binary mass transfer is needed.

%===================================================================================================
\section{Concluding remarks}\label{sec:conclusions}
The observed luminosity extension of RSGs in young open clusters presents a clear challenge to the single-star origin scenario.
In this \textit{Letter}, we present a comprehensive analysis of the luminosity of RSGs in young open clusters (with ages up to 100\,Myr), using state-of-the-art detailed binary-evolution models that incorporate empirical initial binary parameter distributions. These models have previously been employed to successfully explain the observed multiple components of MS stars in young open clusters. 

Our findings reveal that single-star evolution and processes resembling single-star evolution, including pre/non-interaction binaries and pre-MS or ZAMS mergers typically produce RSGs that occupy the lowest luminosity range, spanning approximately 0.14 dex in $\log L$. In contrast, MS mergers result in a significantly broader luminosity spread of up to a factor of 10 (1.0 dex in $\log L$). This luminosity extension remains robust despite uncertainties in the physics of both single and binary star evolution. This predicted luminosity extension aligns well with the observed spread of RSG luminosities in young open clusters, suggesting that the observed high-luminosity red stragglers are strong candidates for being products of MS mergers. 

However, our predictions indicate an overabundance of low-luminosity RSGs from scenarios resembling single-star evolution, which does not match observations. To reconcile this discrepancy, we find that factors such as observational accuracy, RSG variability, and a small age spread of less than approximately 5\,Myr are necessary. These uncertainties, combined with the intrinsic luminosity spread caused by single-star evolution, can produce a total luminosity spread of up to a factor of three (0.5 dex), offering a much better explanation for the observed luminosity distributions of RSGs. Nevertheless, fully addressing this discrepancy will require future studies on accurately measuring the luminosities and variabilities of RSGs in a larger sample of young open clusters, with particular attention to the low-luminosity population.

Our findings suggest that, instead of relying on the previously proposed least-luminous RSG or the average luminosity of the entire RSG population to determine cluster age, the average luminosity of the low- to intermediate-luminosity RSG population (up to a factor of approximately three) serves as a more reliable indicator of cluster age.

Our study offers valuable insights into the formation of RSGs with varying luminosities in young open clusters, paving the way for future observational campaigns. The higher-luminosity RSGs offer a unique opportunity to explore the physics and frequency of binary mergers, whereas detailed investigations of the lower-luminosity population are essential for advancing our understanding of diverse evolutionary pathways, including single-star evolution, stable mass transfer, and post-MS mergers.

%===================================================================================================

\acknowledgments
We sincerely appreciate the valuable comments from the anonymous referee, which have greatly enhanced the clarity of this paper. CW thank Andrea Ercolino, Emanouil Manos Zapartas and Ylva G\"{o}tberg for their insightful discussions and valuable suggestions. 
CW, SJ, and AVG acknowledge funding from the Netherlands Organisation for Scientific Research (NWO), as part of the Vidi research program BinWaves (project number 639.042.728, PI: de Mink). LRP acknowledges support by grants
PID2019-105552RB-C41, PID2022-137779OB-C41 and PID2022-140483NB-C22 funded by
MCIN/AEI/10.13039/501100011033 by "ERDF A way of making
Europe".
EL acknowledges funding by the European Research Council (ERC) under the European Union’s Horizon 2020 research and innovation program (Grant agreement No. 945806) and support by the Klaus Tschira Stiftung.

\newpage

%===================================================================================================
\appendix
\twocolumngrid 
%===================================================================================================

\section{Single-star evolution and comparison with observations}\label{app_sec:3LMC_cluster}
\setcounter{figure}{0}
\renewcommand\thesection{\Alph{section}}
\renewcommand{\thefigure}{\thesection.\arabic{figure}}
\makeatletter
\renewcommand{\theHfigure}{\thesection.\arabic{figure}} % Update hyperref anchors
\makeatother

\begin{figure}[htbp]
%FIG.1
\centering
  \begin{minipage}{0.49\textwidth} 
     \centering
      \includegraphics[width=\linewidth]{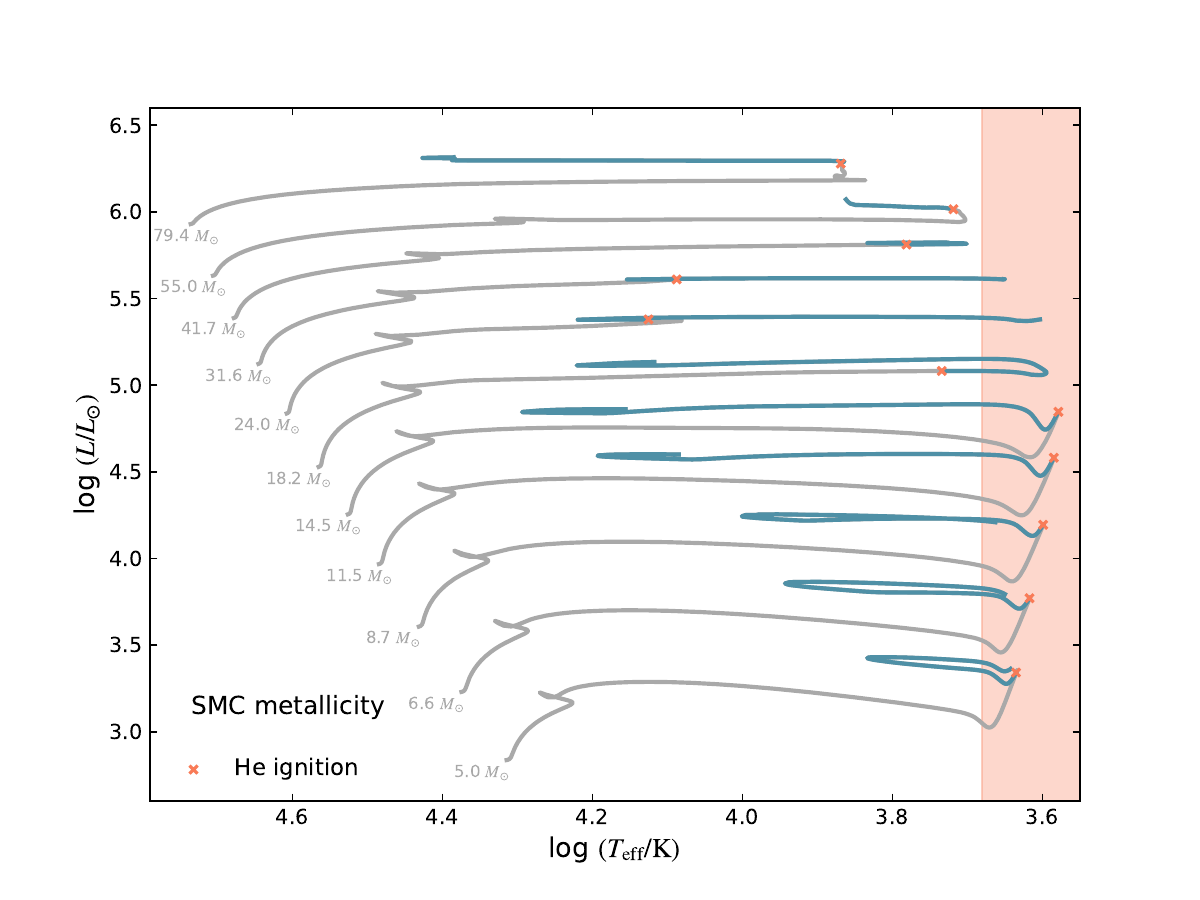}
   \end{minipage}
 \begin{minipage}{0.49\textwidth}         
    \centering
     \includegraphics[width=\linewidth]{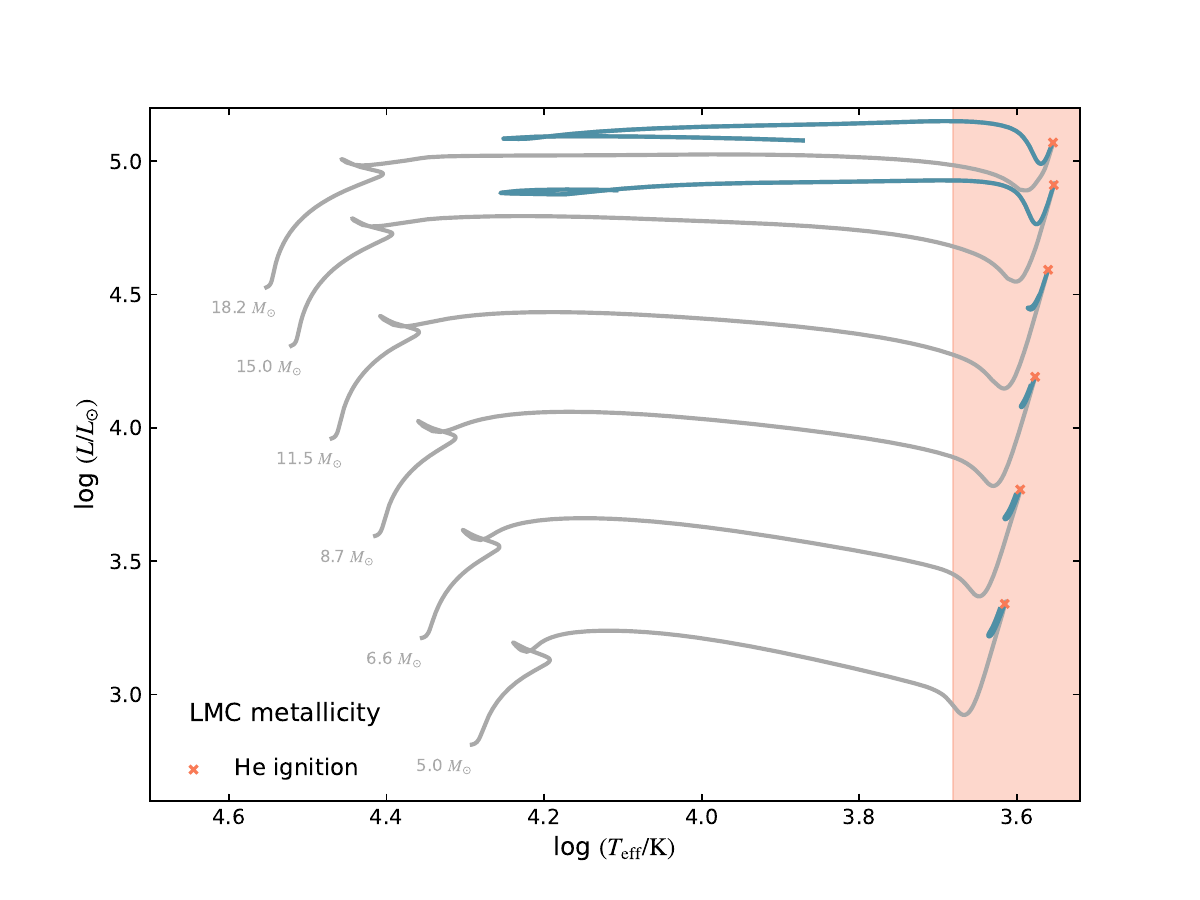}
   \end{minipage}
\caption{Evolution of our non-rotating single-star models with SMC metallicity (upper) and LMC metallicity (lower) in the Hertzsprung-Russell diagram. Grey lines indicate the pre-helium ignition phase, while blue lines represent the central helium burning phase. Orange crosses mark the positions where central helium ignition occurs. The reddish shaded area indicates the region where stars are classified as RSGs in this study.} 
\label{appfig:RSG_single}
\end{figure}

In Figure\,\ref{appfig:RSG_single}, we present the evolution of our SMC and LMC single-star models in the HRD, with central helium ignition and helium burning phases highlighted. The appearance of a blue loop is shown to be sensitive to metallicity. SMC models below approximately 18$\mso$ ignite helium as RSGs and subsequently undergo a blue loop excursion. In contrast, for LMC models, only stars more massive than about 15$\mso$ exhibit a blue loop. For SMC models above 18$\mso$, stars ignite helium at higher temperatures and may or may not become RSGs.

\begin{figure*}[htbp]
%FIG.1
\centering
\includegraphics[width=\linewidth]{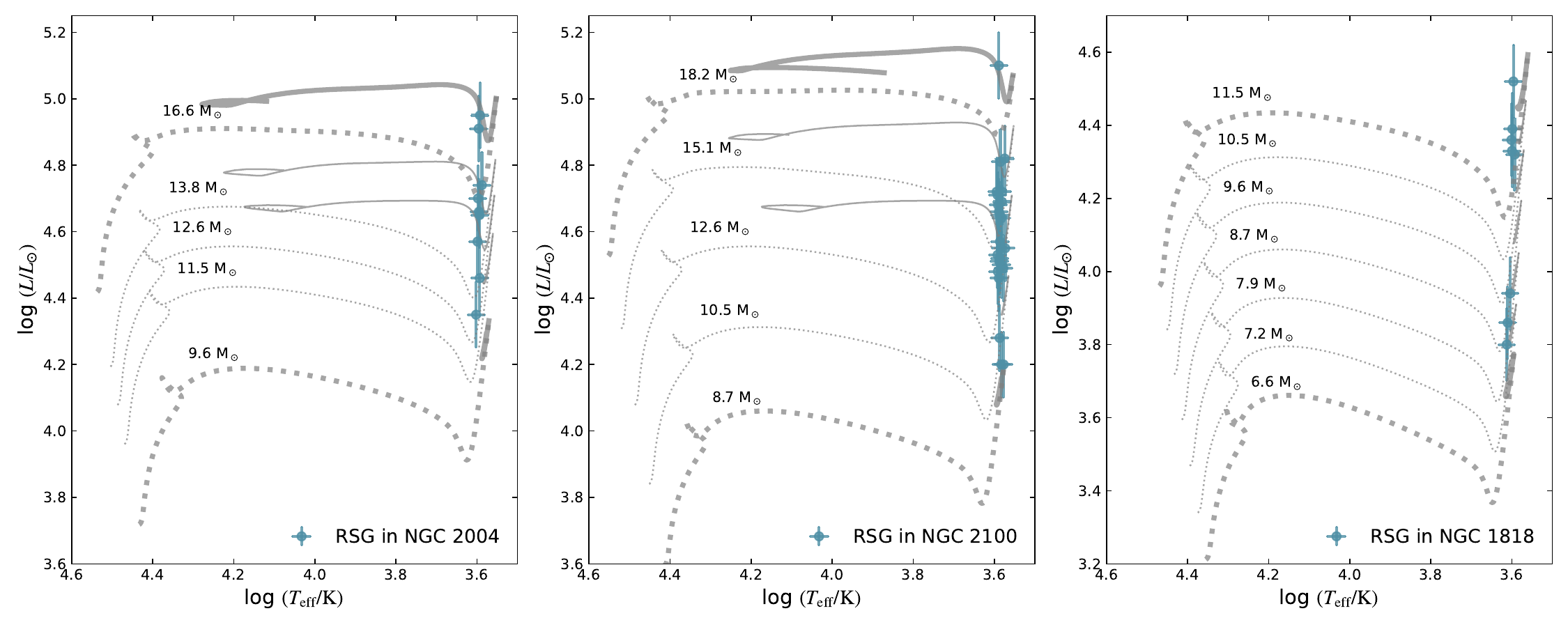}
\caption{Same as Fig.\,\ref{fig:n330_single_track}, but for the comparison of observed RSGs in three LMC clusters, NGC\,2004, NGC\,2100 and NGC\,1818  (\citealt{2016MNRAS.463.1269B,2019MNRAS.486..266B} and N. Britavskiy in private communication), with our LMC single-star models. 
}
\label{appfig:3cluster_single_track}
\end{figure*}

In Figs.\,\ref{appfig:3cluster_single_track}, we compare LMC single-star models with RSGs detected in three young LMC clusters. The same conclusion is achieved that the luminosity ranges of the observed RSGs in these clusters are too large to be explained by single-star evolution. The apparent ages of the least and most luminous RSGs derived from these single-star tracks are 25 and 13\,Myr for NGC\,2004, 30 and 11\,Myr for NGC\,2100, and 48 and 21\,Myr for NGC\,1818, respectively. Here we utilize non-rotating star models. As discussed in \cite{2023A&A...670A..43W}, stellar rotation does not significantly affect the evolution of stars in our models unless they initially rotate at exceptionally high velocities (larger than approximately 70\% of their break-up velocities) due to the assumption of inefficient rotational mixing.

\begin{figure}[htbp]
%FIG.1
\centering
  \begin{minipage}{0.49\textwidth} 
     \centering
      \includegraphics[width=\linewidth]{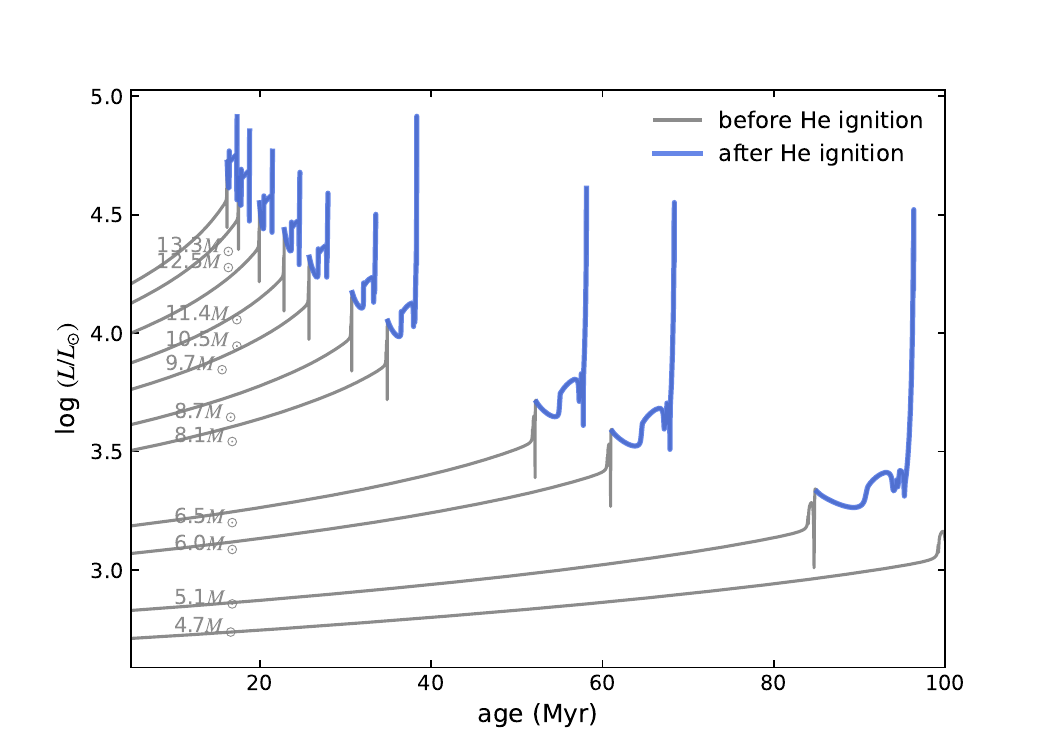}
   \end{minipage}
 \begin{minipage}{0.49\textwidth}         
    \centering
     \includegraphics[width=\linewidth]{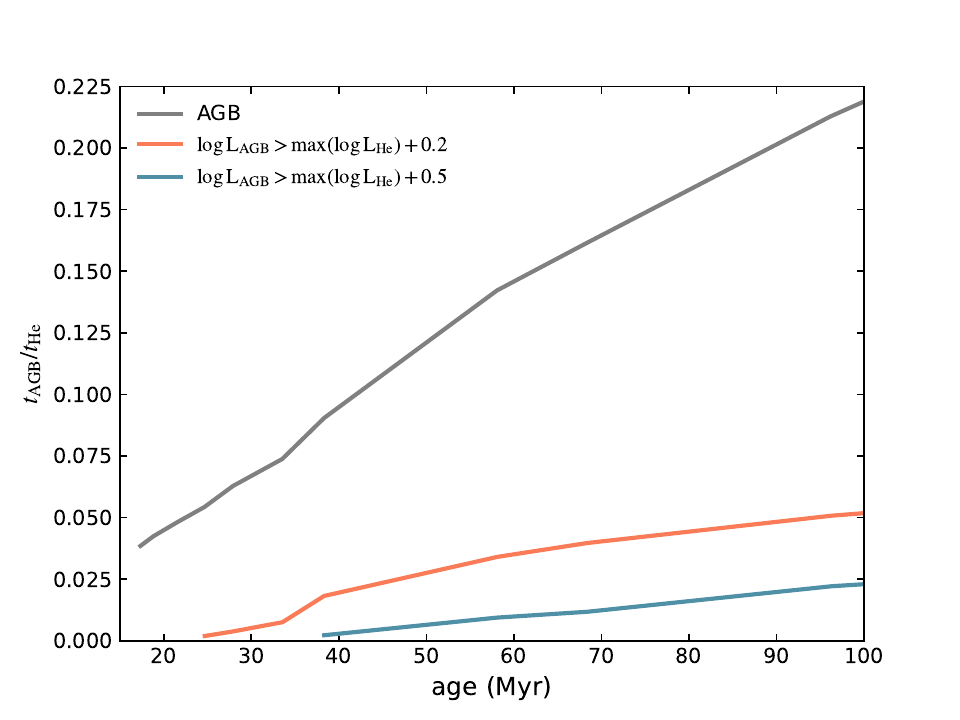}
   \end{minipage}
\caption{Upper: Luminosity evolution of exemplary models in our non-interacting binaries, shown before (grey dashed lines) and after (blue solid lines) central helium exhaustion. Lower: The ratio of the AGB star lifetime to that of their helium-burning counterparts. The grey line represents the entire AGB phase, while the orange and blue lines correspond to AGB stars with luminosities exceeding the maximum luminosity during their helium-burning phase by 0.2 dex and 0.5 dex, respectively.} 
\label{appfig:AGB}
\end{figure}

Our single-star models terminate at the end of central helium exhaustion. Here we also examine post-helium burning evolution using our non-interacting binary models, which are followed until the end of carbon depletion. In Fig.\,\ref{appfig:AGB}, we present the luminosity evolution of these models alongside the lifetime ratio of AGB stars to helium-burning stars. The result show that after helium exhaustion, stars experience a rapid increase in luminosity. Consequently, the contamination of AGB stars within the helium-burning star population remains low, with approximately 4\% in 20\,Myr clusters and up to $\sim$20\% in 100\,Myr clusters. Notably, the fraction of significantly more luminous AGB stars, those exceeding their helium-burning counterparts by 0.5 dex (or 1 dex), is consistently below 2\% (or 0.3\%). These findings suggest that while AGB stars may account for the observed low-luminosity population of RSGs, they are insufficient to explain the high-luminosity population observed in young open clusters.

\begin{figure}[htbp]
%FIG.1
\centering
\includegraphics[width=1.1\linewidth]{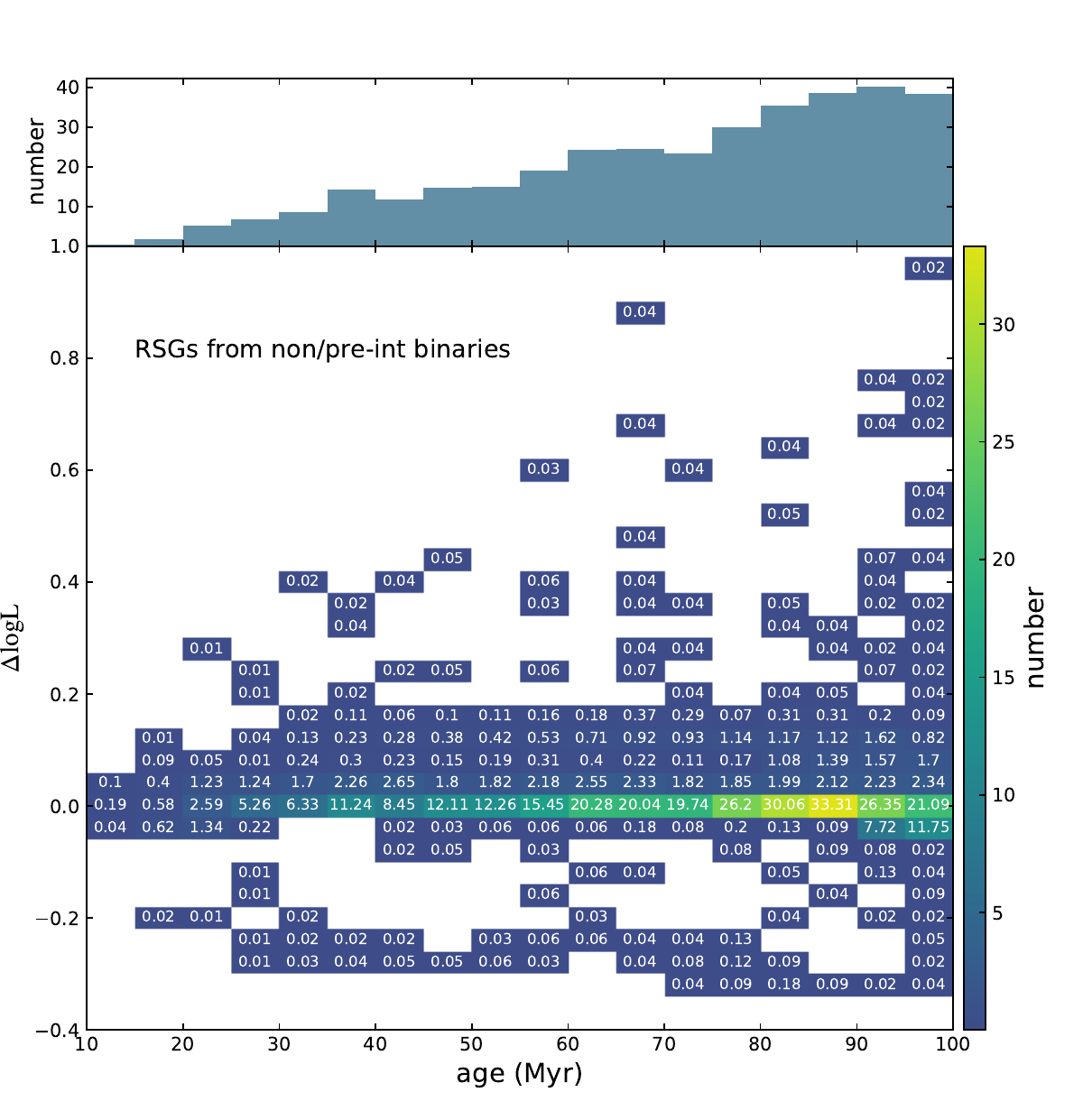}
\caption{Same as Fig.\,\ref{fig:delta_L_accretion}, but for RSGs originating from binaries that have not undergone mass transfer or will never do so.
}
\label{appfig:delta_L_primary}
\end{figure}

In this section, we also present the luminosity extension $\Delta \log L$ of RSGs that evolved from pre/non-interaction binaries that resemble single-star evolution in Fig.\,\ref{appfig:delta_L_primary}. The rare cases of $\Delta \log L \gtrsim 0.1$ are due to contaminations from AGB stars. Whereas the rare cases of $\Delta \log L \lesssim -0.06$ are due to contaminations from pre-helium burning red stars.

\section{Binary evolutionary examples}
\setcounter{figure}{0}
\renewcommand\thesection{\Alph{section}}
\renewcommand{\thefigure}{\thesection.\arabic{figure}}
\makeatletter
\renewcommand{\theHfigure}{\thesection.\arabic{figure}} % Update hyperref anchors
\makeatother

\begin{figure*}[htbp]
\centering
\includegraphics[width=\linewidth]{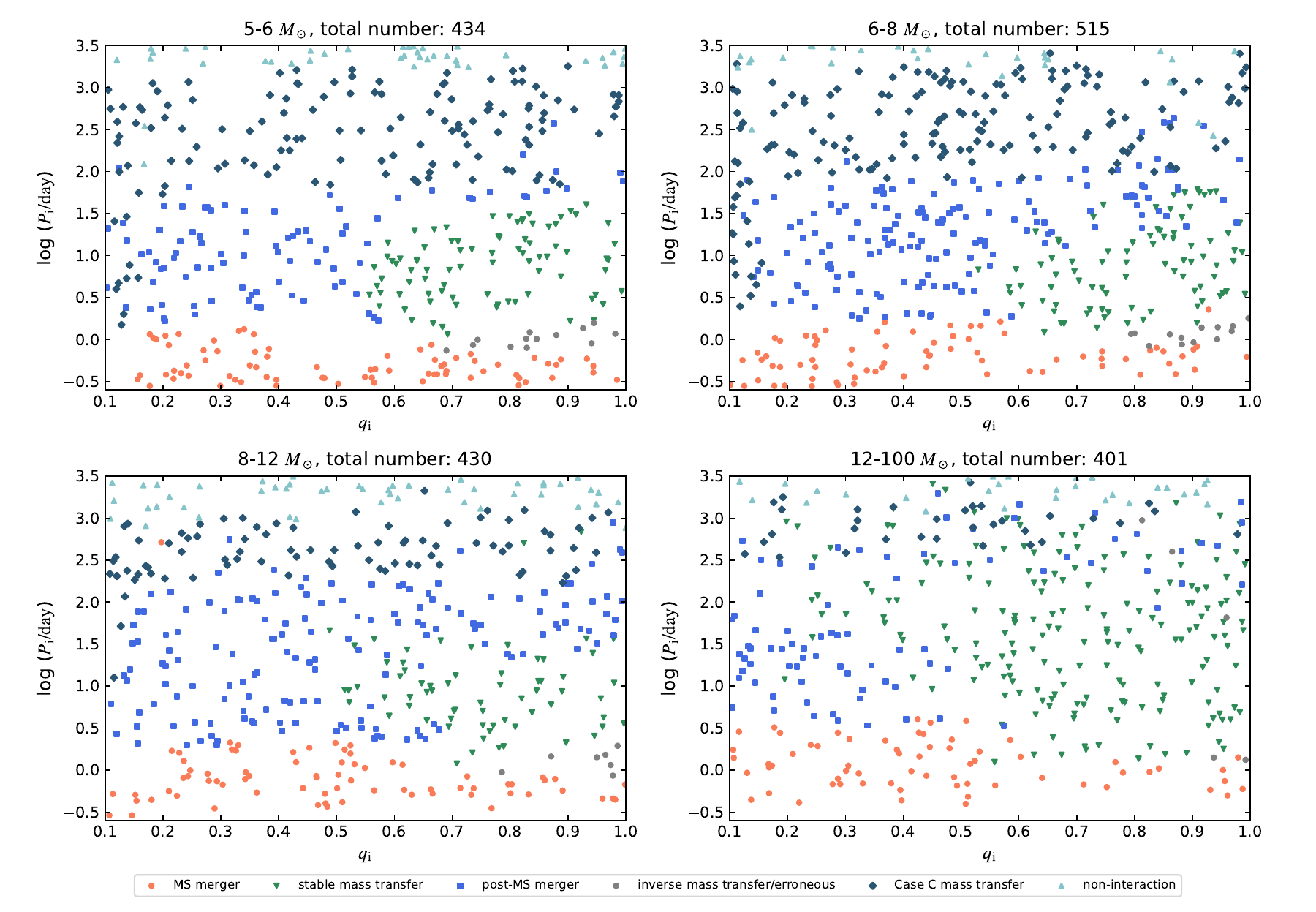}
\caption{Evolution outcomes of our detailed binary models. Each marker represents a binary model with the initial primary mass within the indicated mass range. The initial mass ratio (defined as the primary star’s mass divided by the secondary star’s mass) and orbital periods are shown on the x- and y-axis, respectively. The total number of binaries in each panel is also provided. We identify six distinct evolutionary outcomes for our binary models: main-sequence mergers (orange circles), stable mass transfer (green up-side-down triangles), post-main-sequence mergers (blue squares), inverse mass transfer/errors (grey circles), Case C mass transfer (dark blue diamonds), and wide binaries that do not interaction (light blue triangles). In our simulations, RSGs can originate from main-sequence mergers, stable mass transfer, non-interaction binaries and Case C binaries before mass transfer occurs.}
\label{appfig:outcome}
\end{figure*}

In Fig.\,\ref{appfig:outcome}, we display the evolutionary outcomes of our detailed binary models. We focus on binaries with primary masses above 5$\mso$, as we only consider clusters up to 100\,Myr. For binaries experiencing inverse mass transfer or encountering numerical errors, we stop the calculations as soon as these events occur. Such cases constitute a small fraction (around 2\%) of the models. Inverse mass transfer binaries typically start with high mass ratios, allowing the secondary star’s mass to exceed that of the primary during the first mass transfer phase. In these systems, the secondary star completes hydrogen burning first and subsequently transfers material to the primary. Due to extreme mass ratios, these binaries are expected to merge, and similar to normal Case B mergers, they may primarily produce BSGs/YSGs.

For binaries undergoing Case C mass transfer, we stop calculations once the mass transfer rate reaches the ad-hoc threshold of $10^{-1}\mso \rm{yr}^{-1}$ and assume these systems will merge. Our simulations include the contributions to the RSG population from primary stars prior to the onset of Case C mass transfer. We disregard the possibility that the secondary star could survive common envelope evolution and later become an RSG. This assumption is reasonable, as even if the envelope is successfully ejected, the post-common-envelope orbital separation would likely be too small to permit RSG formation. Instead, the secondary would transfer material to the remnant of the primary before evolving into an RSG.

\begin{figure}[htbp]
%FIG.1
\centering
  \begin{minipage}{0.49\textwidth} 
  \centering
   \includegraphics[width=\linewidth]{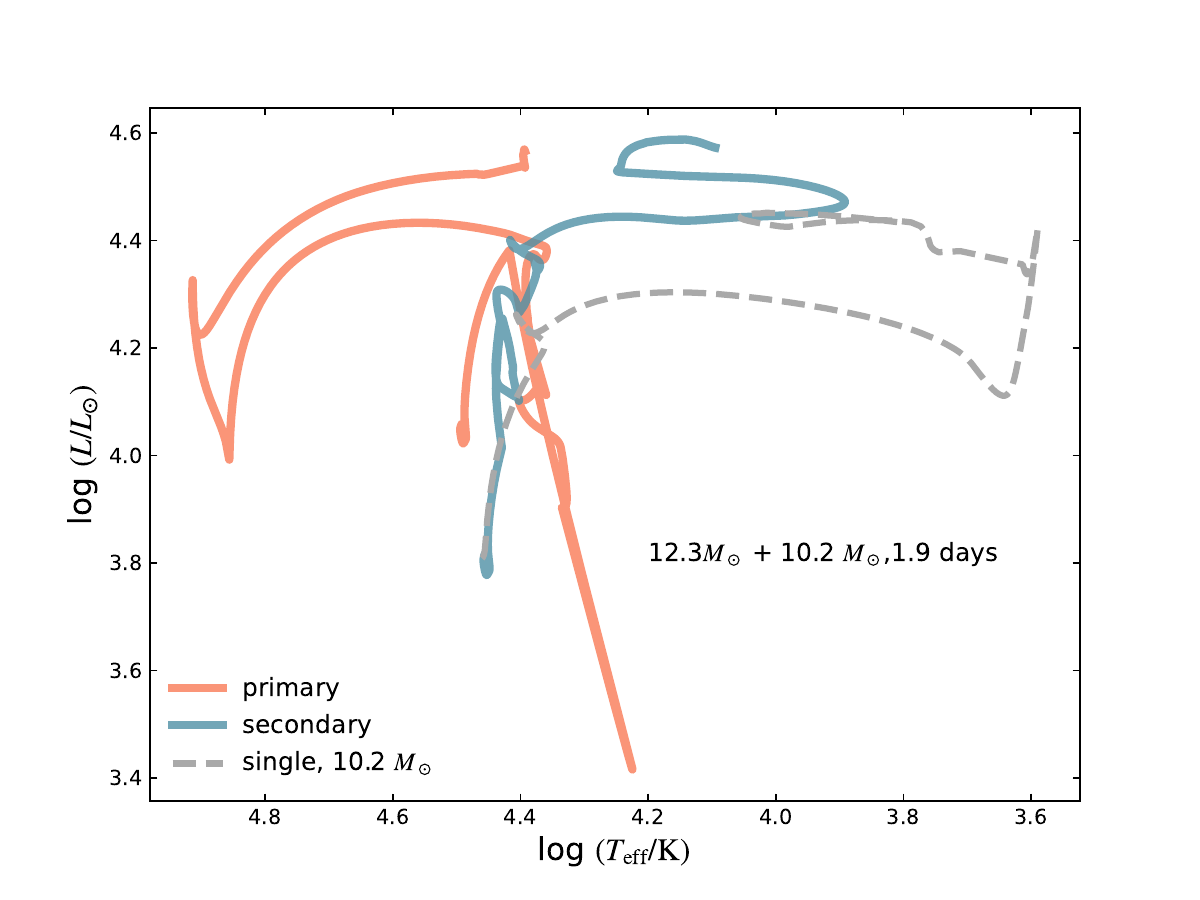}
   \end{minipage}
   \hspace{0pt}
 \begin{minipage}{0.49\textwidth} 
  \centering
   \includegraphics[width=\linewidth]{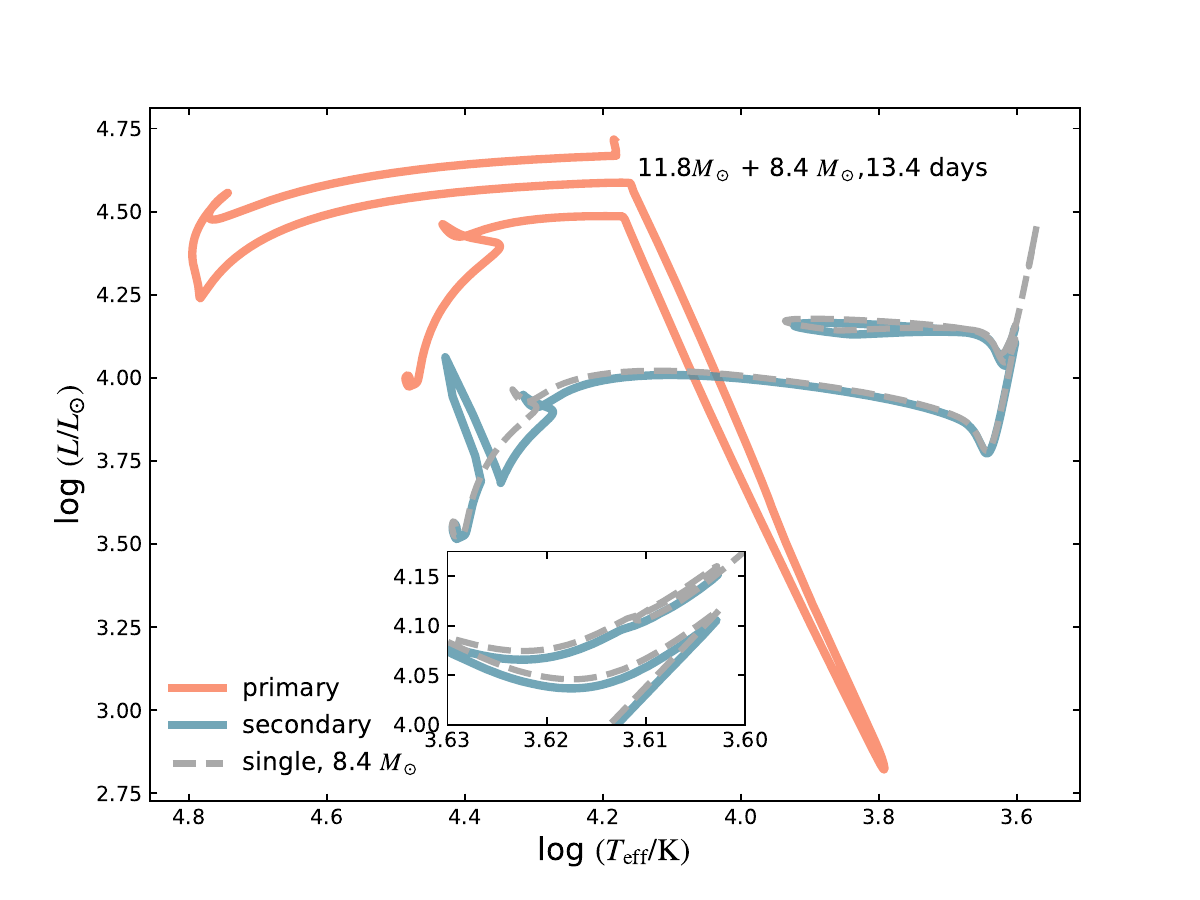} 
   \end{minipage}
   
\caption{Evolution of both components of two binary systems (upper: Case A, lower: Case B) with initial parameters indicated. The orange and blue solid lines depict the evolutionary tracks of the primary and secondary star, respectively. The grey dashed line represents the evolution of a single star model that has the same mass as the initial mass of the secondary star. The inset of the lower panel is a zoom-in of the tip of the supergiant phase.
}
\label{appfig:ex}
\end{figure}

Figure\,\ref{appfig:ex} displays evolution of two example binary models in the HRD, corresponding to Case A mass transfer and Case B mass transfer, respectively. For Case B mass transfer, despite the accretor reaching critical rotation after accretion, its subsequent evolution closely resembles that of a genuine single star with the same total mass. This is because rotationally induced mixing remains negligible due to the presence of a strong chemical gradient in the stellar interior. The upper panel shows that although the accretor in Case A mass transfer accretes more material, it does not contribute to the red straggler population, as it remains blue throughout the entire helium-burning phase.

\section{Definition of Baseline luminosity $L_0$}\label{app_sec:L0}
In this section, we describe how we define the baseline luminosity of RSGs ($L_0$), used to assess the luminosity spread of RSGs in young open clusters. In Figure\,\ref{appfig:RSG_L0}, we show the luminosity evolution over time for our single-star models (see Fig.\,\ref{appfig:RSG_single}). For each mass, we calculate the weighted mean logarithmic luminosity $\log L_0$ during the RSG phase (we ignore the second RSG phase after the blue loop if it exists, as 
its duration is much shorter than the first RSG phase) according to the following expression: 
$$\log L_0 = \frac{\sum_{i=i_\mathrm{start}}^{i=i_\mathrm{end}}\log L_{i}*(t_{i+1}-t_{i})}{\sum_{i=i_\mathrm{start}}^{i=i_\mathrm{end}} (t_{i+1}-t_{i})},$$
where $i_\mathrm{start}$ and $i_\mathrm{end}$ are the indices for the start and end of the RSG phase in MESA history file ($i$ is the model number in MESA history file), and $\log L_{i}$ and $t_{i}$ denote the logarithmic luminosity and age at index $i$. This approach accounts for the time the star spends at each luminosity.
We define the mean age, $\mathrm{age}_0$, of the RSG phase as $(t_{i_\mathrm{end}}-t_{i_\mathrm{start}})/2$. Then we get one data point ($\mathrm{age}_0$, $\log L_0$) for each track in Fig.\,\ref{appfig:RSG_L0}. Using spline interpolation, we obtain the orange line in this figure, representing the evolution of $\log L_0$ over age. This figure shows that $L_0$ closely aligns with the helium ignition luminosity.

The choice of $L_0$ has minimal impact on our conclusions, as single stars exhibit only slight luminosity variation during the RSG phase. Comparing $L_0$ for LMC models (orange line in the lower panel of Fig.\,\ref{appfig:RSG_L0}) and SMC models (black dotted line in the lower panel of Fig.\,\ref{appfig:RSG_L0}), we find that metallicity has a negligible effect on $L_0$. This support our comparison of detailed binary models with SMC metallicity to observed RSGs in LMC clusters (Figures\,\ref{fig:delta_L_merger} and \ref{fig:RSG_number}).

\setcounter{figure}{0}
\renewcommand\thesection{\Alph{section}}
\renewcommand{\thefigure}{\thesection.\arabic{figure}}
\makeatletter
\renewcommand{\theHfigure}{\thesection.\arabic{figure}} % Update hyperref anchors
\makeatother

\begin{figure}[htbp]
\centering
 \vspace{-5pt}  % Adjusts space above the figure
  \begin{minipage}{0.49\textwidth} 
     \centering
      \includegraphics[width=\linewidth]{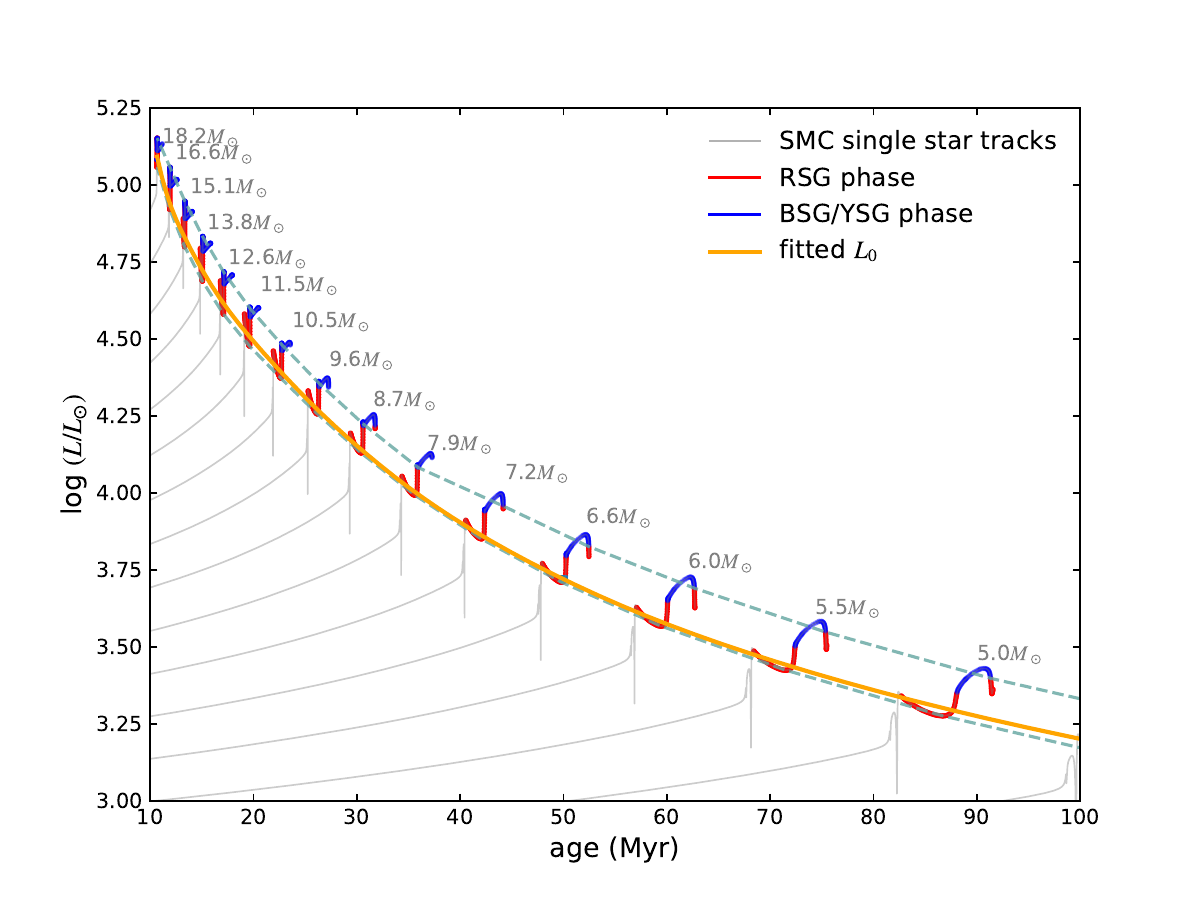}
   \end{minipage}
   \hspace{0pt}
 \begin{minipage}{0.49\textwidth}         
    \centering
     \includegraphics[width=\linewidth]{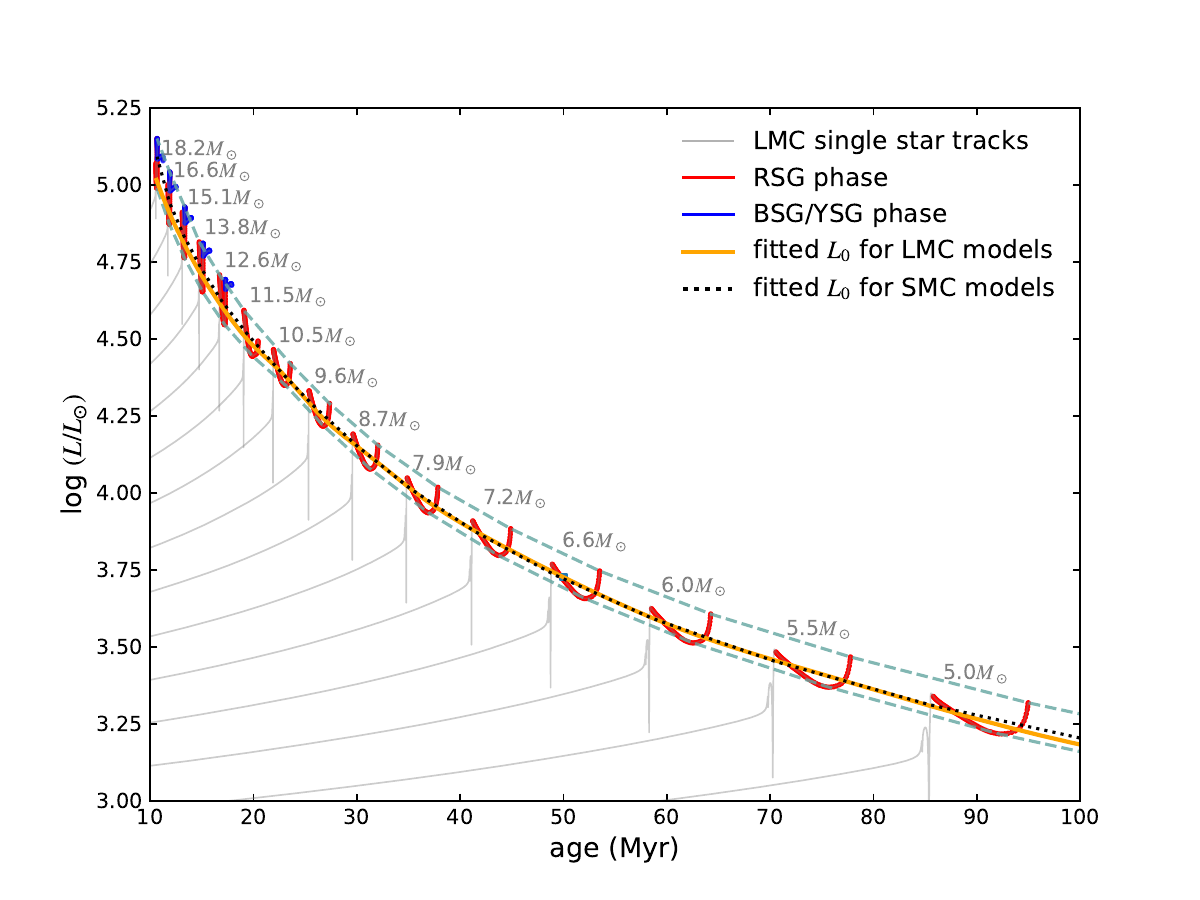}
   \end{minipage}
    \vspace{-5pt}  % Adjusts space above the figure
\caption{Luminosity-age diagram for our SMC (upper) and LMC (lower) single-star models. Thin grey lines show the evolution of stellar luminosity over time for models of various masses, as labeled. The red and blue lines denote the RSG phase and BSG/YSG phase, respectively. Turquoise dashed lines delineate the luminosity range during the helium-burning phase. The orange line represents the evolution of $L_0$ (see text) as a function of age. The black dotted line in the right panel is the same as the orange line in the left panel.
} 
\label{appfig:RSG_L0}
\end{figure}

\section{Semi-analytical method for estimating the upper limit for the RSGs from MS merger products}\label{app_sec:semi_analytical}

To address the problem of random sampling, we include a semi-analytical analysis that establishes an upper limit for the luminosity extension due to MS mergers. We hypothesize that the most massive MS star in a cluster likely results from the merger of two turn-off mass stars. Equal-mass mergers exhibit the strongest rejuvenation effects. These stars are expected to evolve into the most massive RSGs at a slightly later time. We use our single-star models with an initial rotational rate of 55\% of the critical values for merger progenitors and single-star models with a rotational rate of 15\% of the critical values for the merger products.

For a given age $t_\mathrm{merge}$, we first calculate the terminal-age MS (TAMS) mass corresponding to this age and assume that two stars of this mass merge at this age. Using the method described in \citet{2016MNRAS.457.2355S}, we compute the mass of the merger product, $m_\mathrm{merger}$, and its apparent age $t_\mathrm{app}$. Next, through interpolation, we identify the single-star model with an initial mass of $m_\mathrm{merger}$ and determine its maximum luminosity, $L_\mathrm{max}$, as well as the age, $t_\mathrm{max}$, at which this maximum luminosity is reached during the RSG phase. The actual age at which the rejuvenated merger product reaches its maximum luminosity is then given by $t = t_\mathrm{max}-t_\mathrm{app}+t_\mathrm{merge}$. By varying $t_\mathrm{merge}$, we derive the relationship between $t$ and $L_\mathrm{max}$, which is represented by the thick solid grey curve in Panel a of Fig.\,\ref{fig:delta_L_merger}.

This calculation accounts for mass loss during the merger process, which is typically less than 10\% of the total mass of the merger progenitor \citep{2013MNRAS.434.3497G}. If no mass loss is assumed during merger process, $\Delta \log L$ would increase by approximately 0.1 dex.

\section{Cumulative luminosity distribution for the RSGs observed in four clusters}
\setcounter{figure}{0}
\renewcommand\thesection{\Alph{section}}
\renewcommand{\thefigure}{\thesection.\arabic{figure}}
\makeatletter
\renewcommand{\theHfigure}{\thesection.\arabic{figure}} % Update hyperref anchors
\makeatother
In Fig.\,\ref{appfig:ob_LF} we present the cumulative luminosity distribution of RSGs in four young open clusters. Using a threshold of $\Delta \log L$ between 0.4 to 0.5 to identify red stragglers, we find red straggler fractions of 22\% in NGC\,2004, 13\% to 42\% in NGC\,2100, 25\% to 30\% in NGC\,330, and 38\% to 63\% in NGC\,1818.

\begin{figure}[htbp]
\centering
\includegraphics[width=\linewidth]{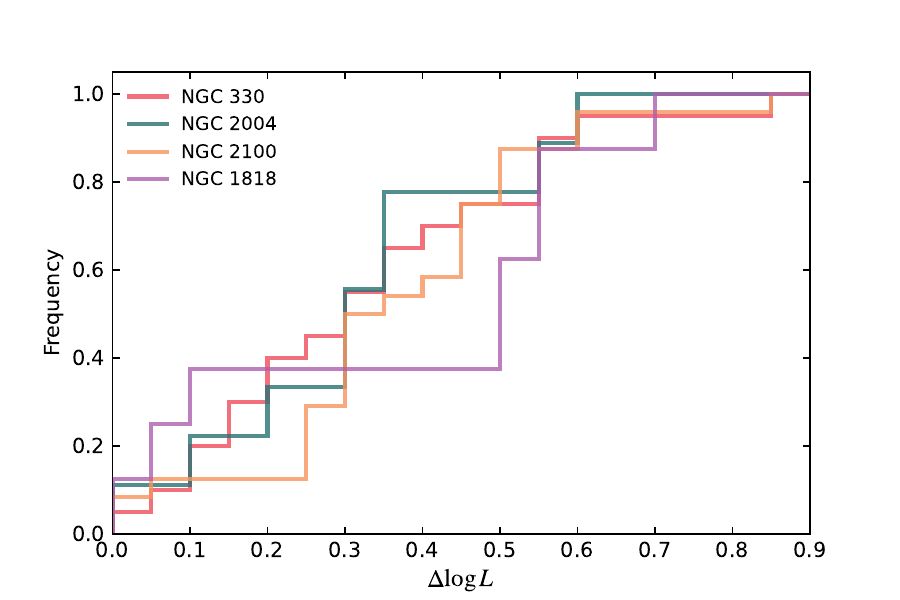}
\caption{Cumulative distribution of $\Delta \log L$ for RSGs in four young open star clusters, as indicated in the legend. For each cluster, $\Delta \log L$ represents the luminosity difference, in logarithmic space, between each RSG and the least-luminous RSG in that cluster. }
\label{appfig:ob_LF}
\end{figure}

\section{Impact of initial binary fraction}\label{app_sec:fb}

\setcounter{figure}{0}
\renewcommand\thesection{\Alph{section}}
\renewcommand{\thefigure}{\thesection.\arabic{figure}}
\makeatletter
\renewcommand{\theHfigure}{\thesection.\arabic{figure}} % Update hyperref anchors
\makeatother
The results and conclusions in our main text are based on the extreme assumption of a binary fraction of 1. To explore the impact of varying initial binary fractions, we conduct a population synthesis simulation using only single-star models. Through a Monte Carlo approach, we generated an initial sample of 3670 single stars (matching the sample size in our binary-based simulation), with initial masses distributed according to the Salpeter IMF within 3 and 100$\mso$ (the same mass range as primary stars in our binary models). We show the number of RSGs as a function of age predicted in this single-star simulation in the left panel of Fig.\,\ref{appfig:fb}. Compared to Fig.\,\ref{fig:RSG_number}, it can be seen that single-star evolution produces more RSGs, as a significant portion of our binary systems undergo post-MS mergers, which likely not substantially contribute to the RSG population. 
%Notably, in single-star evolution, the number of predicted RSGs does not increase with age in constant age bins. This is because the increasing number of lower-mass stars (due to the IMF) is balanced by the narrowing mass ranges that contribute to RSGs in each bin as age progresses.
\balance 

By combining this single-star simulation with our binary star results, which represent an initial binary fraction of 50\%, we calculated the fraction of RSGs originating from binary evolution (represented by the darkest blue bars in the right panel of Fig.\,\ref{appfig:fb}). We then increased the binary fraction to 60\% and 70\% by randomly selecting samples from the single-star simulation. The resulting contributions of binary evolution to RSG formation are shown with lighter blue bars in the right panel of Fig.\,\ref{appfig:fb}.

\begin{figure}[h]
\centering
\includegraphics[width=1.05\linewidth]{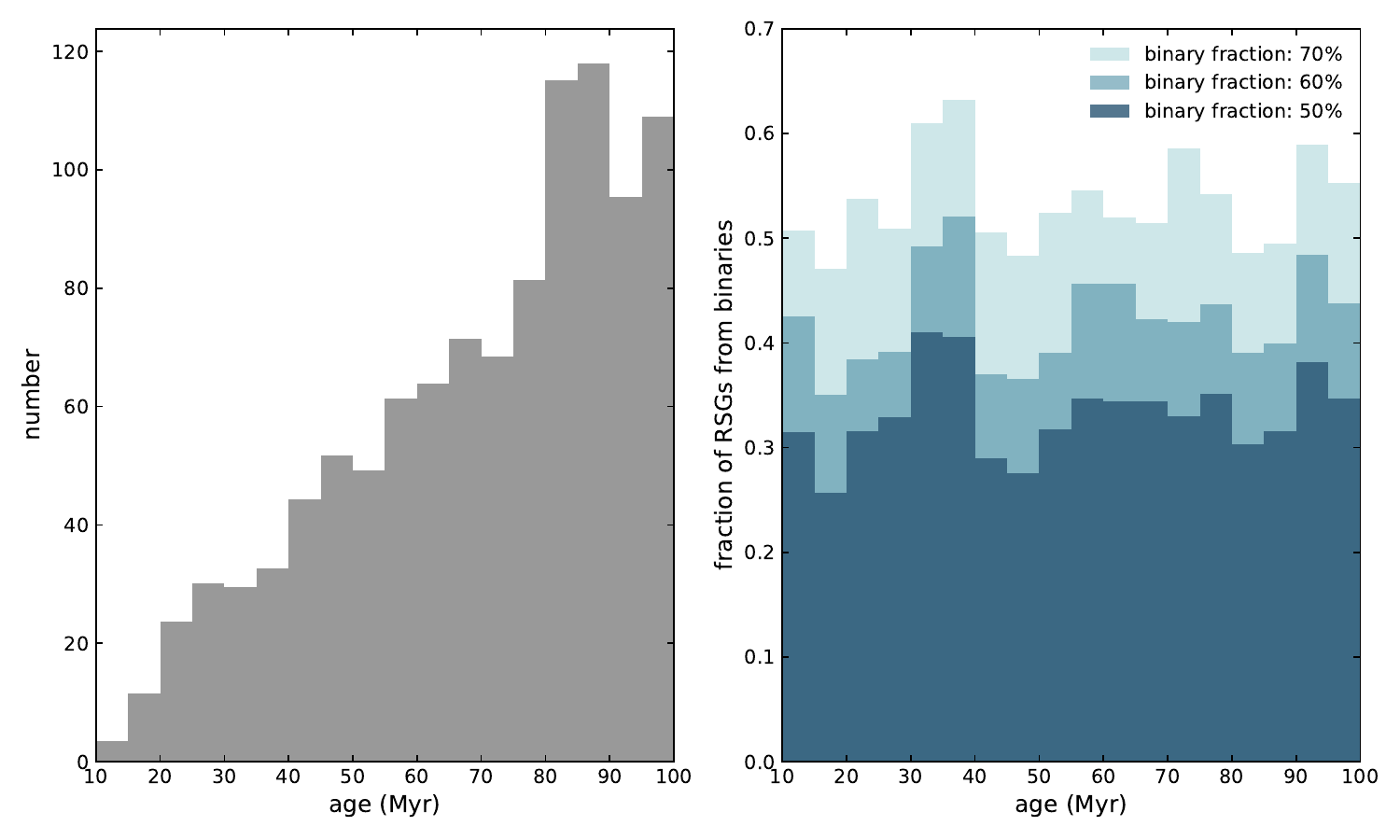}
\caption{Left: Predicted number of RSGs as a function of age from a population synthesis simulation based on 3670 single-star models with initial rotational velocities at 55\% of their critical values. Right: Fraction of RSGs originating from binary evolution, assuming different initial binary fractions as indicated.
}
\label{appfig:fb}
\end{figure}

\section{Impact of supernova kicks and reverse mass transfer}\label{app_sec:SN_kicks}
\setcounter{figure}{0}
\renewcommand\thesection{\Alph{section}}
\renewcommand{\thefigure}{\thesection.\arabic{figure}}
\makeatletter
\renewcommand{\theHfigure}{\thesection.\arabic{figure}} % Update hyperref anchors
\makeatother
As described in the main text, our simulations assume that the secondary star evolves in isolation after the primary star forms a compact object. In this appendix, we investigate the impact of SN kicks that may disrupt binaries and produce runaway RSGs. If a binary remains bound after the SN kick, the secondary star may initiate reverse mass transfer to the primary star before evolving into a RSG. Here we use the prescription in \cite{2024ApJ...975L..20W} to determine the type of compact object formed by the primary star and the SN kicks imparted on them. Specifically, we assume zero-kicks for BHs and a Maxwell-Boltzmann distribution with a root-means-square velocity $\sigma=265 \kms$ for neutron stars (NSs) formed through core-collapse SN \citep{2005MNRAS.360..974H}. For NSs formed from electron capture SN, we assume smaller kicks from a flat distribution between 0 and 50$\kms$ \citep{2002ApJ...574..364P,2004ApJ...612.1044P}. For each binary system, we calculate the probability of the system remaining bound after the SN kicks and use this probability as a weight in our original predictions. For binaries that remain bound (including systems with BH or WD companions), we further compare radius of the RSG formed by the secondary star to the orbital separation at the moment the primary forms a compact object (assuming the orbit remains unchanged thereafter). A RSG is allowed to form only if its radius is smaller than the orbital separation.

\begin{figure}[h]
\centering
\includegraphics[width=\linewidth]{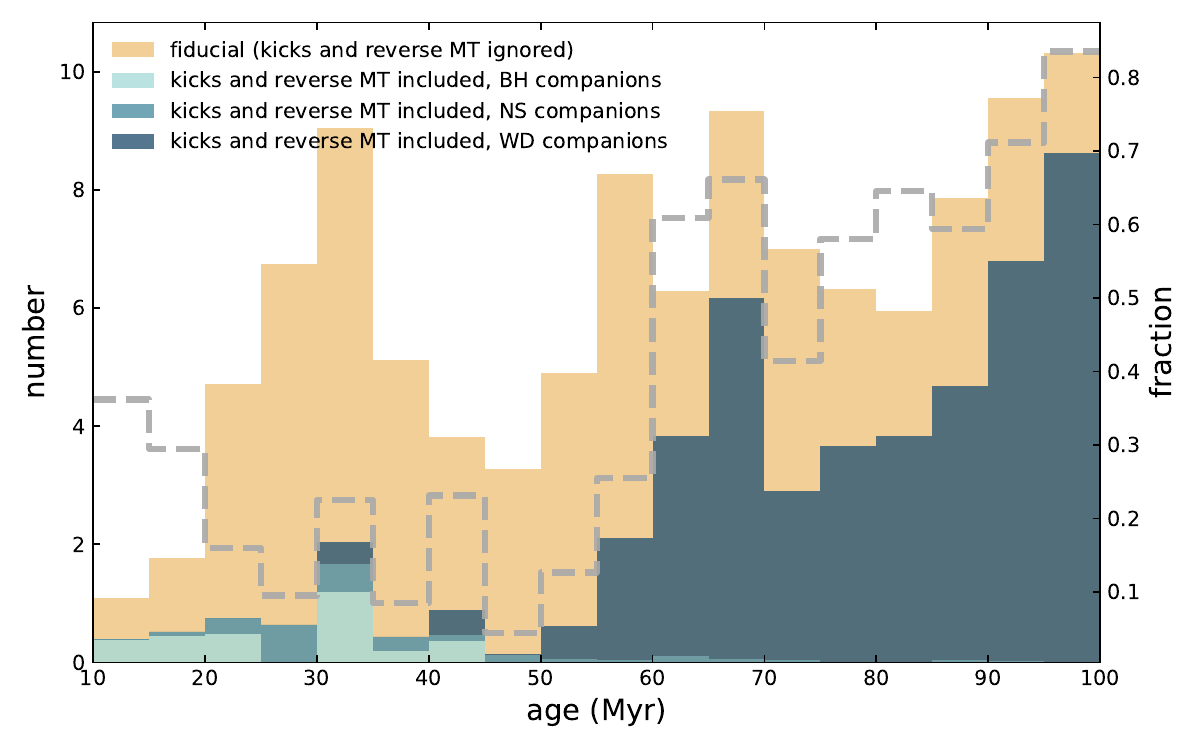}
\caption{The impact of supernova kicks and reverse mass transfer on the formation of RSGs in young open clusters originating from stable mass transfer. The yellow bars represent the predicted number of RSGs in clusters with an initial mass of $1.3\times 10^5\mso$, originating from stable mass transfer in our simulations, as a function of age (the same as the yellow bars in Fig.\,\ref{fig:RSG_number}). The bluish bars show the predicted number of RSGs after accounting for the effects of supernova kicks and reverse mass transfer, with light blue, medium blue and dark blue corresponding to RSGs with black hole, neutron star, and white dwarf companions, respectively. The bluish bars are stacked. The fraction of these newly predicted RSGs, relative to the values from our standard simulations, is indicated by the grey dashed lines and correspond to the scale on the right y-axis.
}
\label{appfig:SN_kicks}
\end{figure}

The results are shown in Fig.\,\ref{appfig:SN_kicks}. Before approximately 50\,Myr, RSGs predominantly have BH or NS companions. For systems where the primary forms a NS, over 90\% are disrupted by SN kicks. For systems where the primary forms a BH, a significant fraction will undergo reverse mass transfer unless the initial orbital period is large (greater than hundreds of days). After accounting for these effects, less than approximately 20\% of RSGs predicted by our original simulations are expected to remain in young open clusters between 20 and 50\,Myr. After 50\,Myr, the primary stars form WDs, and the impact of reverse mass transfer becomes much smaller compared to younger clusters. 

\section{Impact of stellar rotation}\label{app_sec:rotation}
\setcounter{figure}{0}
\renewcommand\thesection{\Alph{section}}
\renewcommand{\thefigure}{\thesection.\arabic{figure}}
\makeatletter
\renewcommand{\theHfigure}{\thesection.\arabic{figure}} % Update hyperref anchors
\makeatother

\begin{figure}[htbp]
%FIG.1
\centering
\includegraphics[width=\linewidth]{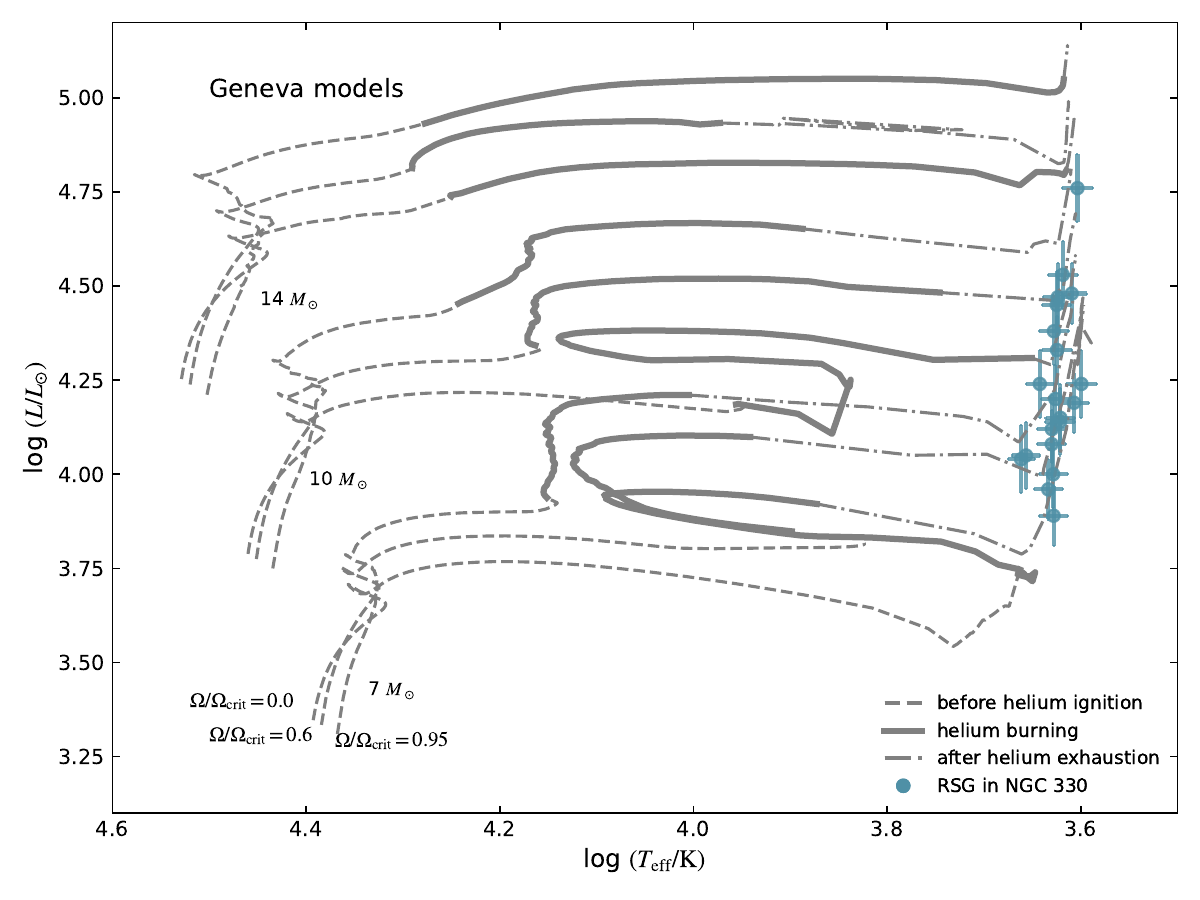}
\caption{Evolution of Geneva single-star models with different initial rotation rates at SMC metallicity in the HRD. The masses are indicated in the figure. Dashed, solid and dash-dotted lines represent phases before central helium ignition, during helium burning and after helium exhaustion, respectively. For each mass, three initial rotational velocities, 0, 60\% and 95\% of the critical angular velocity are included. Observations of RSGs in NGC\,330 are overplotted for comparison. 
}
\label{appfig:Geneva}
\end{figure}

To better understand how rotation impacts the properties of RSGs in young open clusters, we examine Geneva models\footnote{ https://www.unige.ch/sciences/astro/evolution/en/database/syclist/}, which employ more efficient rotationally-induced mixing compared to our models \citep{2013A&A...553A..24G}. Their evolutionary tracks in the HRD are shown in Fig.\,\ref{appfig:Geneva}. It is important to note that the Geneva models, formally labeled as having velocities at 95\% of their critical angular velocities, do not actually rotate near the critical limit. 
As explained in \cite{2023A&A...670A..43W}, due to the initial relaxation phase in the Geneva models, and differences in the definition of critical rotation, this velocity corresponds to only 55\% of critical rotation in our single-star models. 

\begin{figure}[htbp]
%FIG.1
\centering
\includegraphics[width=\linewidth]{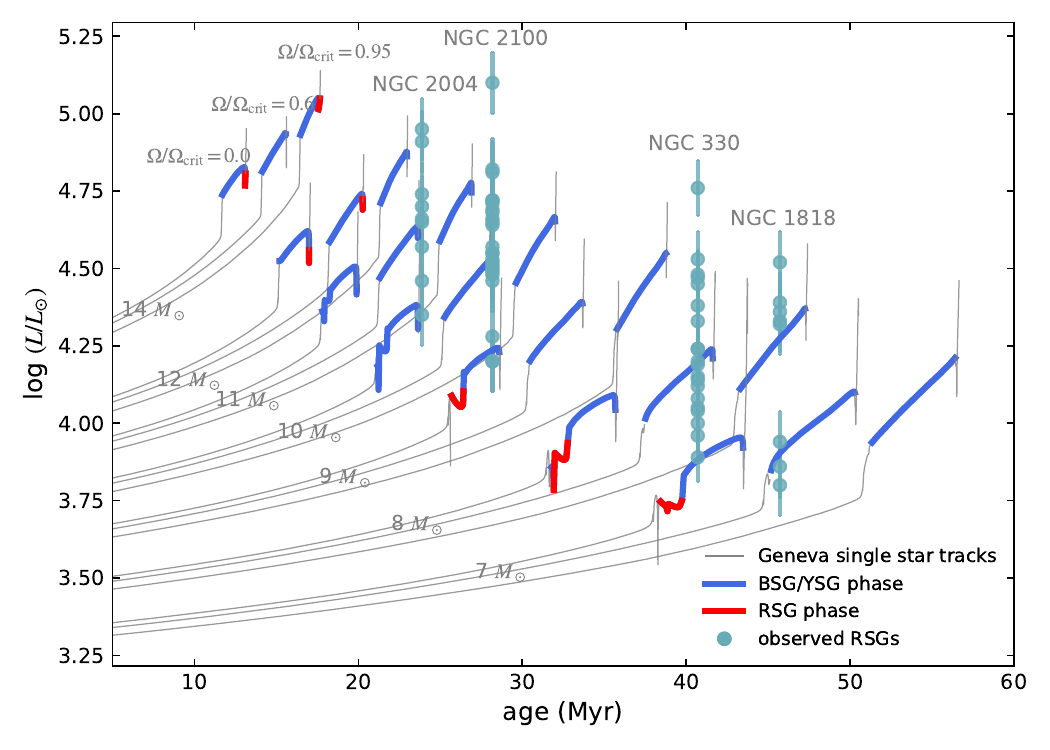}
\caption{Luminosity versus age for Geneva single-star models (thin grey lines) with corresponding masses indicated. For each mass, the three tracks represent initial rotational velocities of 0\%, 60\% and 95\% of the critical angular velocity, shown from left to right. The RSG and BSG/YSG phases are highlighted with thick red and blue lines, respectively. The observed RSGs in the four clusters analyzed in this study are shown as cyan dots with error bars.
}
\label{appfig:Geneva2}
\end{figure}

In Fig.\,\ref{appfig:Geneva}, we see that while fast-rotating models experience significant rejuvenation, which can be seen by their bluer positions near the TAMS and increased luminosity during post-MS evolution, they ignite helium and often complete helium burning as BSGs or YSGs. Therefore, Geneva models cannot explain the observed luminosity spread of RSGs in young open clusters. Similarly, 
Fig.\,\ref{appfig:Geneva2} shows that although rotational mixing leads to a luminosity spread among BSGs, it does not explain the luminosity spread of RSGs. In fact, Fig.\,\ref{appfig:Geneva2} shows that the Geneva models do not predict any RSGs at the ages corresponding to the observed clusters. This is because these models adopt a lower overshooting parameter ($\alpha_\mathrm{ov}=0.1$) and the Schwarzschild criterion to determine the convective boundary, which favor the formation of BSGs over RSGs. To thoroughly investigate whether rotational mixing can lead to a luminosity spread among RSGs, further theoretical studies, incorporating varying parameters for internal mixing, are required.

\section{Impact of real age difference}\label{app_sec:age_diff}

\setcounter{figure}{0}
\renewcommand\thesection{\Alph{section}}
\renewcommand{\thefigure}{\thesection.\arabic{figure}}
\makeatletter
\renewcommand{\theHfigure}{\thesection.\arabic{figure}} % Update hyperref anchors
\makeatother
In this appendix, we explore the impact of potential real age difference on the luminosity of RSGs in young open clusters.
It is important to note that current observations suggest that a significant real age difference among stars in young open clusters is unlikely.
Observations indicate that stars in young open clusters have identical chemical compositions \citep{2014ApJ...793L...6M,2016MNRAS.458.4368M}, suggesting that they belong to the same generation. Previous studies proposed extended star formation as a possible explanation for the extended MS turn off observed in young open clusters \citep{2007MNRAS.379..151M,2014ApJ...797...35G}. \cite{2014ApJ...797...35G} found a mass threshold of approximately $10^{4.8}\mso$ above which clusters can retain stellar ejecta to form a second generation of stars. However, the extended MS turn-off feature has also been observed in clusters with significantly lower masses than this threshold. Notably, two clusters in this study (NGC\,2004 and NGC\,1818) that show a spread luminosity distribution of RSGs, are also measured to be less massive than this threshold. Additionally, \cite{2014MNRAS.443.3594B} examined 13 young open clusters, including NGC\,330 and NGC\,1818, in the Magellanic Clouds and found no evidence of gas or dust within them, which argues against continuous star formation. 

\begin{figure}[htbp]
\centering
\includegraphics[width=\linewidth]{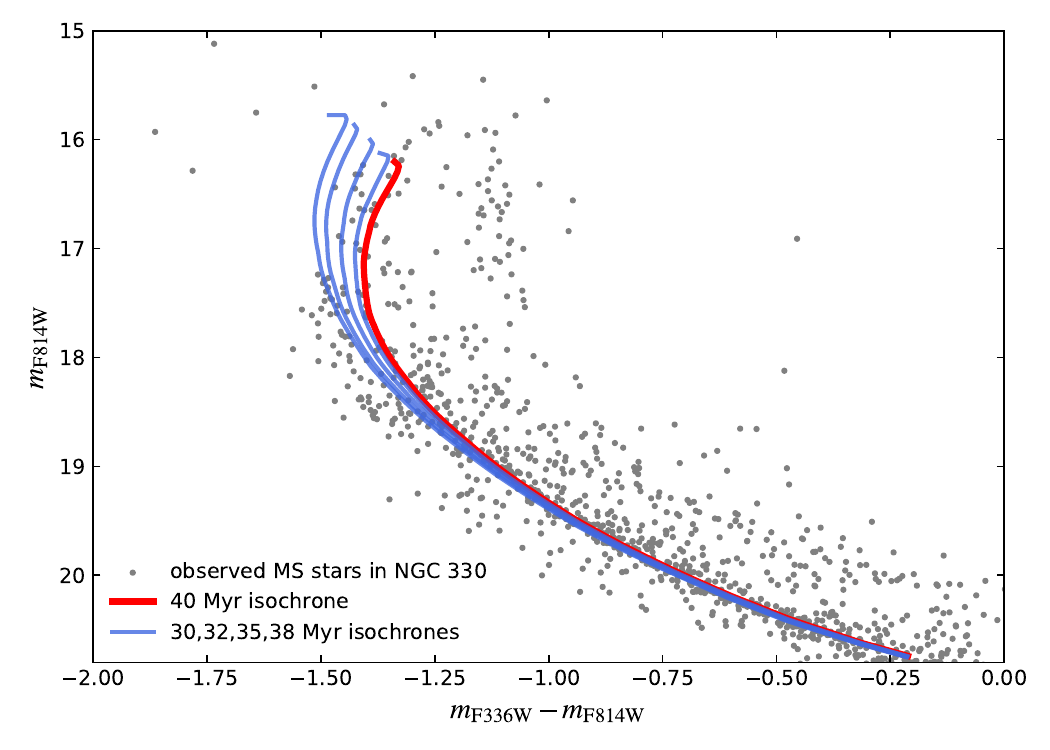}
\caption{Impact of real age difference on the distribution of main-sequence stars in the color-magnitude diagram. The red line represents the 40\,Myr isochrone from our single-star models with initial rotation rates of 55\% of their break-up velocities. The blue lines correspond to the younger isochrones derived from these models, ranging from 30 to 38\,Myr (from left to right). Observed main-sequence stars in the cluster NGC\,330 are overplotted as grey dots. }
\label{appfig:CMD_delta_age_diff}
\end{figure}

In Fig.\,\ref{appfig:CMD_delta_age_diff}, we compare isochrones of different ages, derived from our single-star models, with the observed MS stars in NGC\,330. To transform model temperatures and luminosities into colors and magnitudes, we used the method described in \cite{2022NatAs...6..480W}. For this comparison, we adopted a distance modulus of 18.65 and reddening $E(B-V)=0.07$, such that the 40\,Myr isochrone best fits the majority of MS stars in NGC\,330. Our results show that real age difference cannot account for the blue MS stars fainter than approximately 18.5 magnitude, as unevolved MS stars exhibit nearly identical colors. Meanwhile, real age differences exceeding approximately 5\,Myr cannot explain the observed distribution of MS turn-off stars. This is evident from the lack of stars on the 30 Myr and 32 Myr isochrones above 17.2 magnitude.

\begin{figure}[htbp]
\centering
\includegraphics[width=\linewidth]{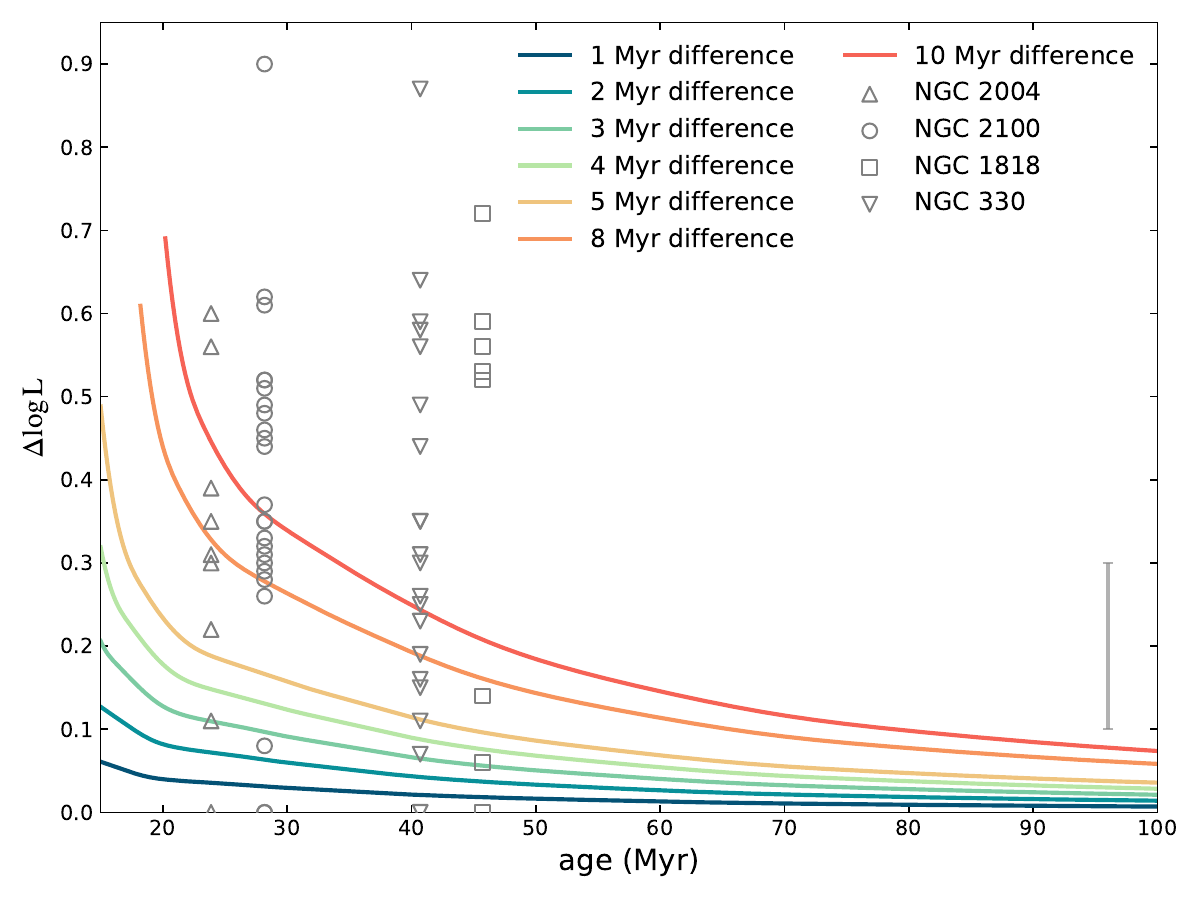}
\caption{Luminosity difference of RSGs as a function of cluster age, caused by real age differences. At a given age, $\Delta \log L$ represents the difference between the baseline luminosity $L_0$ (see Section\,\ref{app_sec:L0}) at that age and the baseline luminosity at an age that is younger by specific values (ranging from 1\,Myr to 10\,Myr), as shown by lines in different colors. The luminosity differences of observed RSGs relative to the least luminous RSG in four young open clusters are overplotted as open markers. Observational uncertainties are illustrated by the grey error bar in the lower right corner.}
\label{appfig:deltaL_age_diff}
\end{figure}

In Fig.\,\ref{appfig:deltaL_age_diff}, we examine the impact of real age difference on the luminosity extension of RSGs in young open clusters. The impact is more significant in younger clusters because at younger ages a given age difference corresponds to a larger mass difference. For example, in a 25\,Myr cluster, a 5\,Myr age difference produces a luminosity difference of approximately 0.18 dex, while a 2\,Myr difference results in a 0.06 dex luminosity spread. In contrast, these values decreases to 0.09 and 0.03 dex, respectively, in a 45\,Myr cluster. 
%Figure\,\ref{appfig:deltaL_age_diff} shows that age differences of up to 5\,Myr can help explain the population of the least luminous RSGs in the four clusters considered in this study.

\bibliographystyle{aasjournal}
\bibliography{RSG_apjl}{}

\begin{thebibliography}{}
\expandafter\ifx\csname natexlab\endcsname\relax\def\natexlab#1{#1}\fi
\providecommand{\url}[1]{\href{#1}{#1}}
\providecommand{\dodoi}[1]{doi:~\href{http://doi.org/#1}{\nolinkurl{#1}}}
\providecommand{\doeprint}[1]{\href{http://ascl.net/#1}{\nolinkurl{http://ascl.net/#1}}}
\providecommand{\doarXiv}[1]{\href{https://arxiv.org/abs/#1}{\nolinkurl{https://arxiv.org/abs/#1}}}

\bibitem[{{Abt}(1983)}]{1983ARA&A..21..343A}
{Abt}, H.~A. 1983, \araa, 21, 343, \dodoi{10.1146/annurev.aa.21.090183.002015}

\bibitem[{{Banyard} {et~al.}(2023){Banyard}, {Mahy}, {Sana}, {Bodensteiner},
  {Villase{\~n}or}, {Sen}, {Langer}, {de Mink}, {Picco}, \&
  {Shenar}}]{2023A&A...674A..60B}
{Banyard}, G., {Mahy}, L., {Sana}, H., {et~al.} 2023, \aap, 674, A60,
  \dodoi{10.1051/0004-6361/202244742}

\bibitem[{{Bastian} \& {Lardo}(2018)}]{2018ARA&A..56...83B}
{Bastian}, N., \& {Lardo}, C. 2018, \araa, 56, 83,
  \dodoi{10.1146/annurev-astro-081817-051839}

\bibitem[{{Bastian} \& {Strader}(2014)}]{2014MNRAS.443.3594B}
{Bastian}, N., \& {Strader}, J. 2014, \mnras, 443, 3594,
  \dodoi{10.1093/mnras/stu1407}

\bibitem[{{Beasor} \& {Davies}(2016)}]{2016MNRAS.463.1269B}
{Beasor}, E.~R., \& {Davies}, B. 2016, \mnras, 463, 1269,
  \dodoi{10.1093/mnras/stw2054}

\bibitem[{{Beasor} {et~al.}(2019){Beasor}, {Davies}, {Smith}, \&
  {Bastian}}]{2019MNRAS.486..266B}
{Beasor}, E.~R., {Davies}, B., {Smith}, N., \& {Bastian}, N. 2019, \mnras, 486,
  266, \dodoi{10.1093/mnras/stz732}

\bibitem[{{Beasor} {et~al.}(2020){Beasor}, {Davies}, {Smith}, {van Loon},
  {Gehrz}, \& {Figer}}]{2020MNRAS.492.5994B}
{Beasor}, E.~R., {Davies}, B., {Smith}, N., {et~al.} 2020, \mnras, 492, 5994,
  \dodoi{10.1093/mnras/staa255}

\bibitem[{{Bellinger} {et~al.}(2024){Bellinger}, {de Mink}, {van Rossem}, \&
  {Justham}}]{2024ApJ...967L..39B}
{Bellinger}, E.~P., {de Mink}, S.~E., {van Rossem}, W.~E., \& {Justham}, S.
  2024, \apjl, 967, L39, \dodoi{10.3847/2041-8213/ad4990}

\bibitem[{{Braun} \& {Langer}(1995)}]{1995A&A...297..483B}
{Braun}, H., \& {Langer}, N. 1995, \aap, 297, 483

\bibitem[{{Britavskiy} {et~al.}(2019){Britavskiy}, {Lennon}, {Patrick},
  {Evans}, {Herrero}, {Langer}, {van Loon}, {Clark}, {Schneider}, {Almeida},
  {Sana}, {de Koter}, \& {Taylor}}]{2019A&A...624A.128B}
{Britavskiy}, N., {Lennon}, D.~J., {Patrick}, L.~R., {et~al.} 2019, \aap, 624,
  A128, \dodoi{10.1051/0004-6361/201834564}

\bibitem[{{Brott} {et~al.}(2011){Brott}, {de Mink}, {Cantiello}, {Langer}, {de
  Koter}, {Evans}, {Hunter}, {Trundle}, \& {Vink}}]{2011A&A...530A.115B}
{Brott}, I., {de Mink}, S.~E., {Cantiello}, M., {et~al.} 2011, \aap, 530, A115,
  \dodoi{10.1051/0004-6361/201016113}

\bibitem[{{Choi} {et~al.}(2016){Choi}, {Dotter}, {Conroy}, {Cantiello},
  {Paxton}, \& {Johnson}}]{2016ApJ...823..102C}
{Choi}, J., {Dotter}, A., {Conroy}, C., {et~al.} 2016, \apj, 823, 102,
  \dodoi{10.3847/0004-637X/823/2/102}

\bibitem[{{Claret} \& {Torres}(2016)}]{2016A&A...592A..15C}
{Claret}, A., \& {Torres}, G. 2016, \aap, 592, A15,
  \dodoi{10.1051/0004-6361/201628779}

\bibitem[{{Davies} {et~al.}(2013){Davies}, {Kudritzki}, {Plez}, {Trager},
  {Lan{\c{c}}on}, {Gazak}, {Bergemann}, {Evans}, \&
  {Chiavassa}}]{2013ApJ...767....3D}
{Davies}, B., {Kudritzki}, R.-P., {Plez}, B., {et~al.} 2013, \apj, 767, 3,
  \dodoi{10.1088/0004-637X/767/1/3}

\bibitem[{{de Mink} {et~al.}(2009){de Mink}, {Cantiello}, {Langer}, {Pols},
  {Brott}, \& {Yoon}}]{2009A&A...497..243D}
{de Mink}, S.~E., {Cantiello}, M., {Langer}, N., {et~al.} 2009, \aap, 497, 243,
  \dodoi{10.1051/0004-6361/200811439}

\bibitem[{{de Mink} {et~al.}(2013){de Mink}, {Langer}, {Izzard}, {Sana}, \& {de
  Koter}}]{2013ApJ...764..166D}
{de Mink}, S.~E., {Langer}, N., {Izzard}, R.~G., {Sana}, H., \& {de Koter}, A.
  2013, \apj, 764, 166, \dodoi{10.1088/0004-637X/764/2/166}

\bibitem[{{Drout} {et~al.}(2009){Drout}, {Massey}, {Meynet}, {Tokarz}, \&
  {Caldwell}}]{2009ApJ...703..441D}
{Drout}, M.~R., {Massey}, P., {Meynet}, G., {Tokarz}, S., \& {Caldwell}, N.
  2009, \apj, 703, 441, \dodoi{10.1088/0004-637X/703/1/441}

\bibitem[{{Ekstr{\"o}m} {et~al.}(2012){Ekstr{\"o}m}, {Georgy}, {Eggenberger},
  {Meynet}, {Mowlavi}, {Wyttenbach}, {Granada}, {Decressin}, {Hirschi},
  {Frischknecht}, {Charbonnel}, \& {Maeder}}]{2012A&A...537A.146E}
{Ekstr{\"o}m}, S., {Georgy}, C., {Eggenberger}, P., {et~al.} 2012, \aap, 537,
  A146, \dodoi{10.1051/0004-6361/201117751}

\bibitem[{{Eldridge} {et~al.}(2020){Eldridge}, {Beasor}, \&
  {Britavskiy}}]{2020MNRAS.495L.102E}
{Eldridge}, J.~J., {Beasor}, E.~R., \& {Britavskiy}, N. 2020, \mnras, 495,
  L102, \dodoi{10.1093/mnrasl/slaa067}

\bibitem[{{Eldridge} {et~al.}(2008){Eldridge}, {Izzard}, \&
  {Tout}}]{2008MNRAS.384.1109E}
{Eldridge}, J.~J., {Izzard}, R.~G., \& {Tout}, C.~A. 2008, \mnras, 384, 1109,
  \dodoi{10.1111/j.1365-2966.2007.12738.x}

\bibitem[{{Farrell} {et~al.}(2019){Farrell}, {Groh}, {Meynet}, {Kudritzki},
  {Eldridge}, {Georgy}, {Ekstr{\"o}m}, \& {Yoon}}]{2019A&A...621A..22F}
{Farrell}, E.~J., {Groh}, J.~H., {Meynet}, G., {et~al.} 2019, \aap, 621, A22,
  \dodoi{10.1051/0004-6361/201833657}

\bibitem[{{Georgy} {et~al.}(2013){Georgy}, {Ekstr{\"o}m}, {Granada}, {Meynet},
  {Mowlavi}, {Eggenberger}, \& {Maeder}}]{2013A&A...553A..24G}
{Georgy}, C., {Ekstr{\"o}m}, S., {Granada}, A., {et~al.} 2013, \aap, 553, A24,
  \dodoi{10.1051/0004-6361/201220558}

\bibitem[{{Glebbeek} {et~al.}(2013){Glebbeek}, {Gaburov}, {Portegies Zwart}, \&
  {Pols}}]{2013MNRAS.434.3497G}
{Glebbeek}, E., {Gaburov}, E., {Portegies Zwart}, S., \& {Pols}, O.~R. 2013,
  \mnras, 434, 3497, \dodoi{10.1093/mnras/stt1268}

\bibitem[{{Goudfrooij} {et~al.}(2014){Goudfrooij}, {Girardi},
  {Kozhurina-Platais}, {Kalirai}, {Platais}, {Puzia}, {Correnti}, {Bressan},
  {Chandar}, {Kerber}, {Marigo}, \& {Rubele}}]{2014ApJ...797...35G}
{Goudfrooij}, P., {Girardi}, L., {Kozhurina-Platais}, V., {et~al.} 2014, \apj,
  797, 35, \dodoi{10.1088/0004-637X/797/1/35}

\bibitem[{{Gratton} {et~al.}(2012){Gratton}, {Carretta}, \&
  {Bragaglia}}]{2012A&ARv..20...50G}
{Gratton}, R.~G., {Carretta}, E., \& {Bragaglia}, A. 2012, \aapr, 20, 50,
  \dodoi{10.1007/s00159-012-0050-3}

\bibitem[{{Guinan} {et~al.}(2020){Guinan}, {Wasatonic}, {Calderwood}, \&
  {Carona}}]{2020ATel13512....1G}
{Guinan}, E., {Wasatonic}, R., {Calderwood}, T., \& {Carona}, D. 2020, The
  Astronomer's Telegram, 13512, 1

\bibitem[{{Haberl} \& {Sturm}(2016)}]{2016A&A...586A..81H}
{Haberl}, F., \& {Sturm}, R. 2016, \aap, 586, A81,
  \dodoi{10.1051/0004-6361/201527326}

\bibitem[{{Hastings} {et~al.}(2021){Hastings}, {Langer}, {Wang},
  {Schootemeijer}, \& {Milone}}]{2021A&A...653A.144H}
{Hastings}, B., {Langer}, N., {Wang}, C., {Schootemeijer}, A., \& {Milone},
  A.~P. 2021, \aap, 653, A144, \dodoi{10.1051/0004-6361/202141269}

\bibitem[{{Hellings}(1983)}]{1983Ap&SS..96...37H}
{Hellings}, P. 1983, \apss, 96, 37, \dodoi{10.1007/BF00661941}

\bibitem[{{Hobbs} {et~al.}(2005){Hobbs}, {Lorimer}, {Lyne}, \&
  {Kramer}}]{2005MNRAS.360..974H}
{Hobbs}, G., {Lorimer}, D.~R., {Lyne}, A.~G., \& {Kramer}, M. 2005, \mnras,
  360, 974, \dodoi{10.1111/j.1365-2966.2005.09087.x}

\bibitem[{{Justham} {et~al.}(2014){Justham}, {Podsiadlowski}, \&
  {Vink}}]{2014ApJ...796..121J}
{Justham}, S., {Podsiadlowski}, P., \& {Vink}, J.~S. 2014, \apj, 796, 121,
  \dodoi{10.1088/0004-637X/796/2/121}

\bibitem[{{Kee} {et~al.}(2021){Kee}, {Sundqvist}, {Decin}, {de Koter}, \&
  {Sana}}]{2021A&A...646A.180K}
{Kee}, N.~D., {Sundqvist}, J.~O., {Decin}, L., {de Koter}, A., \& {Sana}, H.
  2021, \aap, 646, A180, \dodoi{10.1051/0004-6361/202039224}

\bibitem[{{Kippenhahn} \& {Weigert}(1967)}]{1967ZA.....65..251K}
{Kippenhahn}, R., \& {Weigert}, A. 1967, \zap, 65, 251

\bibitem[{{Kiss} {et~al.}(2006){Kiss}, {Szab{\'o}}, \&
  {Bedding}}]{2006MNRAS.372.1721K}
{Kiss}, L.~L., {Szab{\'o}}, G.~M., \& {Bedding}, T.~R. 2006, \mnras, 372, 1721,
  \dodoi{10.1111/j.1365-2966.2006.10973.x}

\bibitem[{{Korntreff} {et~al.}(2012){Korntreff}, {Kaczmarek}, \&
  {Pfalzner}}]{2012A&A...543A.126K}
{Korntreff}, C., {Kaczmarek}, T., \& {Pfalzner}, S. 2012, \aap, 543, A126,
  \dodoi{10.1051/0004-6361/201118019}

\bibitem[{{Kozai}(1962)}]{1962AJ.....67..591K}
{Kozai}, Y. 1962, \aj, 67, 591, \dodoi{10.1086/108790}

\bibitem[{{Li} {et~al.}(2017){Li}, {de Grijs}, {Deng}, \&
  {Milone}}]{2017ApJ...844..119L}
{Li}, C., {de Grijs}, R., {Deng}, L., \& {Milone}, A.~P. 2017, \apj, 844, 119,
  \dodoi{10.3847/1538-4357/aa7b36}

\bibitem[{{Lidov}(1962)}]{1962P&SS....9..719L}
{Lidov}, M.~L. 1962, \planss, 9, 719, \dodoi{10.1016/0032-0633(62)90129-0}

\bibitem[{{Mackey} \& {Broby Nielsen}(2007)}]{2007MNRAS.379..151M}
{Mackey}, A.~D., \& {Broby Nielsen}, P. 2007, \mnras, 379, 151,
  \dodoi{10.1111/j.1365-2966.2007.11915.x}

\bibitem[{{Maeder} \& {Meynet}(2000)}]{Maeder2000}
{Maeder}, A., \& {Meynet}, G. 2000, \araa, 38, 143,
  \dodoi{10.1146/annurev.astro.38.1.143}

\bibitem[{{Massey} {et~al.}(2009){Massey}, {Silva}, {Levesque}, {Plez},
  {Olsen}, {Clayton}, {Meynet}, \& {Maeder}}]{2009ApJ...703..420M}
{Massey}, P., {Silva}, D.~R., {Levesque}, E.~M., {et~al.} 2009, \apj, 703, 420,
  \dodoi{10.1088/0004-637X/703/1/420}

\bibitem[{{McLaughlin} \& {van der Marel}(2005)}]{2005ApJS..161..304M}
{McLaughlin}, D.~E., \& {van der Marel}, R.~P. 2005, \apjs, 161, 304,
  \dodoi{10.1086/497429}

\bibitem[{{Milone} {et~al.}(2009){Milone}, {Bedin}, {Piotto}, \&
  {Anderson}}]{2009A&A...497..755M}
{Milone}, A.~P., {Bedin}, L.~R., {Piotto}, G., \& {Anderson}, J. 2009, \aap,
  497, 755, \dodoi{10.1051/0004-6361/200810870}

\bibitem[{{Milone} {et~al.}(2016){Milone}, {Marino}, {D'Antona}, {Bedin}, {Da
  Costa}, {Jerjen}, \& {Mackey}}]{2016MNRAS.458.4368M}
{Milone}, A.~P., {Marino}, A.~F., {D'Antona}, F., {et~al.} 2016, \mnras, 458,
  4368, \dodoi{10.1093/mnras/stw608}

\bibitem[{{Milone} {et~al.}(2015){Milone}, {Bedin}, {Piotto}, {Marino},
  {Cassisi}, {Bellini}, {Jerjen}, {Pietrinferni}, {Aparicio}, \&
  {Rich}}]{2015MNRAS.450.3750M}
{Milone}, A.~P., {Bedin}, L.~R., {Piotto}, G., {et~al.} 2015, \mnras, 450,
  3750, \dodoi{10.1093/mnras/stv829}

\bibitem[{{Milone} {et~al.}(2018){Milone}, {Marino}, {Di Criscienzo},
  {D'Antona}, {Bedin}, {Da Costa}, {Piotto}, {Tailo}, {Dotter}, {Angeloni},
  {Anderson}, {Jerjen}, {Li}, {Dupree}, {Granata}, {Lagioia}, {Mackey},
  {Nardiello}, \& {Vesperini}}]{2018MNRAS.477.2640M}
{Milone}, A.~P., {Marino}, A.~F., {Di Criscienzo}, M., {et~al.} 2018, \mnras,
  477, 2640, \dodoi{10.1093/mnras/sty661}

\bibitem[{{Milone} {et~al.}(2023){Milone}, {Cordoni}, {Marino}, {D'Antona},
  {Bellini}, {Di Criscienzo}, {Dondoglio}, {Lagioia}, {Langer}, {Legnardi},
  {Libralato}, {Baumgardt}, {Bettinelli}, {Cavecchi}, {de Grijs}, {Deng},
  {Hastings}, {Li}, {Mohandasan}, {Renzini}, {Vesperini}, {Wang}, {Ziliotto},
  {Carlos}, {Costa}, {Dell'Agli}, {Di Stefano}, {Jang}, {Martorano}, {Simioni},
  {Tailo}, \& {Ventura}}]{2023A&A...672A.161M}
{Milone}, A.~P., {Cordoni}, G., {Marino}, A.~F., {et~al.} 2023, \aap, 672,
  A161, \dodoi{10.1051/0004-6361/202244798}

\bibitem[{{Mucciarelli} {et~al.}(2014){Mucciarelli}, {Dalessandro}, {Ferraro},
  {Origlia}, \& {Lanzoni}}]{2014ApJ...793L...6M}
{Mucciarelli}, A., {Dalessandro}, E., {Ferraro}, F.~R., {Origlia}, L., \&
  {Lanzoni}, B. 2014, \apjl, 793, L6, \dodoi{10.1088/2041-8205/793/1/L6}

\bibitem[{{Narloch} {et~al.}(2021){Narloch}, {Pietrzy{\'n}ski}, {Gieren},
  {Piatti}, {G{\'o}rski}, {Karczmarek}, {Graczyk}, {Suchomska}, {Zgirski},
  {Wielg{\'o}rski}, {Pilecki}, {Taormina}, {Ka{\l}uszy{\'n}ski}, {Pych},
  {Hajdu}, \& {Rojas Garc{\'\i}a}}]{2021A&A...647A.135N}
{Narloch}, W., {Pietrzy{\'n}ski}, G., {Gieren}, W., {et~al.} 2021, \aap, 647,
  A135, \dodoi{10.1051/0004-6361/202039623}

\bibitem[{{Neo} {et~al.}(1977){Neo}, {Miyaji}, {Nomoto}, \&
  {Sugimoto}}]{1977PASJ...29..249N}
{Neo}, S., {Miyaji}, S., {Nomoto}, K., \& {Sugimoto}, D. 1977, \pasj, 29, 249

\bibitem[{{Niederhofer} {et~al.}(2015){Niederhofer}, {Hilker}, {Bastian}, \&
  {Silva-Villa}}]{2015A&A...575A..62N}
{Niederhofer}, F., {Hilker}, M., {Bastian}, N., \& {Silva-Villa}, E. 2015,
  \aap, 575, A62, \dodoi{10.1051/0004-6361/201424455}

\bibitem[{{Offner} {et~al.}(2023){Offner}, {Moe}, {Kratter}, {Sadavoy},
  {Jensen}, \& {Tobin}}]{2023ASPC..534..275O}
{Offner}, S.~S.~R., {Moe}, M., {Kratter}, K.~M., {et~al.} 2023, in Astronomical
  Society of the Pacific Conference Series, Vol. 534, Protostars and Planets
  VII, ed. S.~{Inutsuka}, Y.~{Aikawa}, T.~{Muto}, K.~{Tomida}, \& M.~{Tamura},
  275, \dodoi{10.48550/arXiv.2203.10066}

\bibitem[{{{\"O}pik}(1924)}]{1924PTarO..25f...1O}
{{\"O}pik}, E. 1924, Publications of the Tartu Astrofizica Observatory, 25, 1

\bibitem[{{Packet}(1981)}]{1981A&A...102...17P}
{Packet}, W. 1981, \aap, 102, 17

\bibitem[{{Patrick} {et~al.}(2016){Patrick}, {Evans}, {Davies}, {Kudritzki},
  {H{\'e}nault-Brunet}, {Bastian}, {Lapenna}, \&
  {Bergemann}}]{2016MNRAS.458.3968P}
{Patrick}, L.~R., {Evans}, C.~J., {Davies}, B., {et~al.} 2016, \mnras, 458,
  3968, \dodoi{10.1093/mnras/stw561}

\bibitem[{{Patrick} {et~al.}(2024){Patrick}, {Lennon}, {Schootemeijer},
  {Bianchi}, {Negueruela}, {Langer}, {Thilker}, \&
  {Dorda}}]{2024arXiv241218554P}
{Patrick}, L.~R., {Lennon}, D.~J., {Schootemeijer}, A., {et~al.} 2024, arXiv
  e-prints, arXiv:2412.18554, \dodoi{10.48550/arXiv.2412.18554}

\bibitem[{{Patrick} {et~al.}(2020){Patrick}, {Lennon}, {Evans}, {Sana},
  {Bodensteiner}, {Britavskiy}, {Dorda}, {Herrero}, {Negueruela}, \& {de
  Koter}}]{2020A&A...635A..29P}
{Patrick}, L.~R., {Lennon}, D.~J., {Evans}, C.~J., {et~al.} 2020, \aap, 635,
  A29, \dodoi{10.1051/0004-6361/201936741}

\bibitem[{{Paxton} {et~al.}(2011){Paxton}, {Bildsten}, {Dotter}, {Herwig},
  {Lesaffre}, \& {Timmes}}]{Paxton2011}
{Paxton}, B., {Bildsten}, L., {Dotter}, A., {et~al.} 2011, \apjs, 192, 3,
  \dodoi{10.1088/0067-0049/192/1/3}

\bibitem[{{Paxton} {et~al.}(2013){Paxton}, {Cantiello}, {Arras}, {Bildsten},
  {Brown}, {Dotter}, {Mankovich}, {Montgomery}, {Stello}, {Timmes}, \&
  {Townsend}}]{Paxton2013}
{Paxton}, B., {Cantiello}, M., {Arras}, P., {et~al.} 2013, \apjs, 208, 4,
  \dodoi{10.1088/0067-0049/208/1/4}

\bibitem[{{Paxton} {et~al.}(2015){Paxton}, {Marchant}, {Schwab}, {Bauer},
  {Bildsten}, {Cantiello}, {Dessart}, {Farmer}, {Hu}, {Langer}, {Townsend},
  {Townsley}, \& {Timmes}}]{Paxton2015}
{Paxton}, B., {Marchant}, P., {Schwab}, J., {et~al.} 2015, \apjs, 220, 15,
  \dodoi{10.1088/0067-0049/220/1/15}

\bibitem[{{Paxton} {et~al.}(2018){Paxton}, {Schwab}, {Bauer}, {Bildsten},
  {Blinnikov}, {Duffell}, {Farmer}, {Goldberg}, {Marchant}, {Sorokina},
  {Thoul}, {Townsend}, \& {Timmes}}]{2018ApJS..234...34P}
{Paxton}, B., {Schwab}, J., {Bauer}, E.~B., {et~al.} 2018, \apjs, 234, 34,
  \dodoi{10.3847/1538-4365/aaa5a8}

\bibitem[{{Pfahl} {et~al.}(2002){Pfahl}, {Rappaport}, {Podsiadlowski}, \&
  {Spruit}}]{2002ApJ...574..364P}
{Pfahl}, E., {Rappaport}, S., {Podsiadlowski}, P., \& {Spruit}, H. 2002, \apj,
  574, 364, \dodoi{10.1086/340794}

\bibitem[{{Podsiadlowski}(1992)}]{1992PASP..104..717P}
{Podsiadlowski}, P. 1992, \pasp, 104, 717, \dodoi{10.1086/133043}

\bibitem[{{Podsiadlowski} {et~al.}(2004){Podsiadlowski}, {Langer},
  {Poelarends}, {Rappaport}, {Heger}, \& {Pfahl}}]{2004ApJ...612.1044P}
{Podsiadlowski}, P., {Langer}, N., {Poelarends}, A.~J.~T., {et~al.} 2004, \apj,
  612, 1044, \dodoi{10.1086/421713}

\bibitem[{{Portegies Zwart} {et~al.}(2010){Portegies Zwart}, {McMillan}, \&
  {Gieles}}]{2010ARA&A..48..431P}
{Portegies Zwart}, S.~F., {McMillan}, S. L.~W., \& {Gieles}, M. 2010, \araa,
  48, 431, \dodoi{10.1146/annurev-astro-081309-130834}

\bibitem[{{Rocha} {et~al.}(2024){Rocha}, {Kalogera}, {Doctor}, {Andrews},
  {Sun}, {Gossage}, {Bavera}, {Fragos}, {Kovlakas}, {Kruckow}, {Misra},
  {Srivastava}, {Xing}, \& {Zapartas}}]{2024ApJ...971..133R}
{Rocha}, K.~A., {Kalogera}, V., {Doctor}, Z., {et~al.} 2024, \apj, 971, 133,
  \dodoi{10.3847/1538-4357/ad5955}

\bibitem[{{Salpeter}(1955)}]{Salpeter1955}
{Salpeter}, E.~E. 1955, \apj, 121, 161, \dodoi{10.1086/145971}

\bibitem[{{Sana} {et~al.}(2012){Sana}, {de Mink}, {de Koter}, {Langer},
  {Evans}, {Gieles}, {Gosset}, {Izzard}, {Le Bouquin}, \&
  {Schneider}}]{2012Sci...337..444S}
{Sana}, H., {de Mink}, S.~E., {de Koter}, A., {et~al.} 2012, Science, 337, 444,
  \dodoi{10.1126/science.1223344}

\bibitem[{{Schneider} {et~al.}(2016){Schneider}, {Podsiadlowski}, {Langer},
  {Castro}, \& {Fossati}}]{2016MNRAS.457.2355S}
{Schneider}, F.~R.~N., {Podsiadlowski}, P., {Langer}, N., {Castro}, N., \&
  {Fossati}, L. 2016, \mnras, 457, 2355, \dodoi{10.1093/mnras/stw148}

\bibitem[{{Schneider} {et~al.}(2024){Schneider}, {Podsiadlowski}, \&
  {Laplace}}]{2024A&A...686A..45S}
{Schneider}, F.~R.~N., {Podsiadlowski}, P., \& {Laplace}, E. 2024, \aap, 686,
  A45, \dodoi{10.1051/0004-6361/202347854}

\bibitem[{{Schootemeijer} {et~al.}(2019){Schootemeijer}, {Langer}, {Grin}, \&
  {Wang}}]{2019A&A...625A.132S}
{Schootemeijer}, A., {Langer}, N., {Grin}, N.~J., \& {Wang}, C. 2019, \aap,
  625, A132, \dodoi{10.1051/0004-6361/201935046}

\bibitem[{{Sen} {et~al.}(2022){Sen}, {Langer}, {Marchant}, {Menon}, {de Mink},
  {Schootemeijer}, {Sch{\"u}rmann}, {Mahy}, {Hastings}, {Nathaniel}, {Sana},
  {Wang}, \& {Xu}}]{2022A&A...659A..98S}
{Sen}, K., {Langer}, N., {Marchant}, P., {et~al.} 2022, \aap, 659, A98,
  \dodoi{10.1051/0004-6361/202142574}

\bibitem[{{Shao} \& {Li}(2014)}]{2014ApJ...796...37S}
{Shao}, Y., \& {Li}, X.-D. 2014, \apj, 796, 37,
  \dodoi{10.1088/0004-637X/796/1/37}

\bibitem[{{Shenar} {et~al.}(2022){Shenar}, {Sana}, {Mahy}, {Ma{\'\i}z
  Apell{\'a}niz}, {Crowther}, {Gromadzki}, {Herrero}, {Langer}, {Marchant},
  {Schneider}, {Sen}, {Soszy{\'n}ski}, \& {Toonen}}]{2022A&A...665A.148S}
{Shenar}, T., {Sana}, H., {Mahy}, L., {et~al.} 2022, \aap, 665, A148,
  \dodoi{10.1051/0004-6361/202244245}

\bibitem[{{Smith}(2014)}]{2014ARA&A..52..487S}
{Smith}, N. 2014, \araa, 52, 487, \dodoi{10.1146/annurev-astro-081913-040025}

\bibitem[{{Soraisam} {et~al.}(2018){Soraisam}, {Bildsten}, {Drout}, {Bauer},
  {Gilfanov}, {Kupfer}, {Laher}, {Masci}, {Prince}, {Kulkarni}, {Matheson}, \&
  {Saha}}]{2018ApJ...859...73S}
{Soraisam}, M.~D., {Bildsten}, L., {Drout}, M.~R., {et~al.} 2018, \apj, 859,
  73, \dodoi{10.3847/1538-4357/aabc59}

\bibitem[{{Tokovinin} \& {Moe}(2020)}]{2020MNRAS.491.5158T}
{Tokovinin}, A., \& {Moe}, M. 2020, \mnras, 491, 5158,
  \dodoi{10.1093/mnras/stz3299}

\bibitem[{{Vinciguerra} {et~al.}(2020){Vinciguerra}, {Neijssel},
  {Vigna-G{\'o}mez}, {Mandel}, {Podsiadlowski}, {Maccarone}, {Nicholl},
  {Kingdon}, {Perry}, \& {Salemi}}]{2020MNRAS.498.4705V}
{Vinciguerra}, S., {Neijssel}, C.~J., {Vigna-G{\'o}mez}, A., {et~al.} 2020,
  \mnras, 498, 4705, \dodoi{10.1093/mnras/staa2177}

\bibitem[{{von Zeipel}(1910)}]{1910AN....183..345V}
{von Zeipel}, H. 1910, Astronomische Nachrichten, 183, 345,
  \dodoi{10.1002/asna.19091832202}

\bibitem[{{Wang} {et~al.}(2020){Wang}, {Langer}, {Schootemeijer}, {Castro},
  {Adscheid}, {Marchant}, \& {Hastings}}]{2020ApJ...888L..12W}
{Wang}, C., {Langer}, N., {Schootemeijer}, A., {et~al.} 2020, \apjl, 888, L12,
  \dodoi{10.3847/2041-8213/ab6171}

\bibitem[{{Wang} {et~al.}(2022){Wang}, {Langer}, {Schootemeijer}, {Milone},
  {Hastings}, {Xu}, {Bodensteiner}, {Sana}, {Castro}, {Lennon}, {Marchant},
  {Koter}, \& {Mink}}]{2022NatAs...6..480W}
---. 2022, Nature Astronomy, 6, 480, \dodoi{10.1038/s41550-021-01597-5}

\bibitem[{{Wang} {et~al.}(2023){Wang}, {Hastings}, {Schootemeijer}, {Langer},
  {de Mink}, {Bodensteiner}, {Milone}, {Justham}, \&
  {Marchant}}]{2023A&A...670A..43W}
{Wang}, C., {Hastings}, B., {Schootemeijer}, A., {et~al.} 2023, \aap, 670, A43,
  \dodoi{10.1051/0004-6361/202245413}

\bibitem[{{Wang} {et~al.}(2024){Wang}, {Bodensteiner}, {Xu}, {de Mink},
  {Langer}, {Laplace}, {Vigna-G{\'o}mez}, {Justham}, {Klencki}, {Olejak},
  {Valli}, \& {Schootemeijer}}]{2024ApJ...975L..20W}
{Wang}, C., {Bodensteiner}, J., {Xu}, X.-T., {et~al.} 2024, \apjl, 975, L20,
  \dodoi{10.3847/2041-8213/ad86b7}

\bibitem[{{Wood} {et~al.}(1983){Wood}, {Bessell}, \&
  {Fox}}]{1983ApJ...272...99W}
{Wood}, P.~R., {Bessell}, M.~S., \& {Fox}, M.~W. 1983, \apj, 272, 99,
  \dodoi{10.1086/161265}

\bibitem[{{Yang} {et~al.}(2018){Yang}, {Bonanos}, {Jiang}, {Gao}, {Xue},
  {Wang}, {Lam}, {Spetsieri}, {Ren}, \& {Gavras}}]{2018A&A...616A.175Y}
{Yang}, M., {Bonanos}, A.~Z., {Jiang}, B.-W., {et~al.} 2018, \aap, 616, A175,
  \dodoi{10.1051/0004-6361/201832833}

\bibitem[{{Yang} {et~al.}(2023){Yang}, {Bonanos}, {Jiang}, {Zapartas}, {Gao},
  {Ren}, {Lam}, {Wang}, {Maravelias}, {Gavras}, {Wang}, {Chen}, {Tramper}, {de
  Wit}, {Chen}, {Wen}, {Liu}, {Tian}, {Antoniadis}, \&
  {Luo}}]{2023A&A...676A..84Y}
{Yang}, M., {Bonanos}, A.~Z., {Jiang}, B., {et~al.} 2023, \aap, 676, A84,
  \dodoi{10.1051/0004-6361/202244770}

\bibitem[{{Yoon} \& {Cantiello}(2010)}]{2010ApJ...717L..62Y}
{Yoon}, S.-C., \& {Cantiello}, M. 2010, \apjl, 717, L62,
  \dodoi{10.1088/2041-8205/717/1/L62}

\bibitem[{{Zapartas} {et~al.}(2019){Zapartas}, {de Mink}, {Justham}, {Smith},
  {de Koter}, {Renzo}, {Arcavi}, {Farmer}, {G{\"o}tberg}, \&
  {Toonen}}]{2019A&A...631A...5Z}
{Zapartas}, E., {de Mink}, S.~E., {Justham}, S., {et~al.} 2019, \aap, 631, A5,
  \dodoi{10.1051/0004-6361/201935854}

\bibitem[{{Zapartas} {et~al.}(2024){Zapartas}, {de Wit}, {Antoniadis},
  {Mu{\~n}oz-Sanchez}, {Souropanis}, {Bonanos}, {Maravelias}, {Kovlakas},
  {Kruckow}, {Fragos}, {Andrews}, {Bavera}, {Briel}, {Gossage}, {Kasdagli},
  {Rocha}, {Sun}, {Srivastava}, \& {Xing}}]{2024arXiv241007335Z}
{Zapartas}, E., {de Wit}, S., {Antoniadis}, K., {et~al.} 2024, arXiv e-prints,
  arXiv:2410.07335, \dodoi{10.48550/arXiv.2410.07335}

\end{thebibliography}



\end{document}